\pdfoutput=1
\documentclass[
  reprint,
  superscriptaddress,
  floatfix,
  citeautoscript,
  amsmath,
  amssymb,
  prapplied,          
  longbibliography   
]{revtex4-2}

\usepackage[T1]{fontenc}
\usepackage[utf8]{inputenc}
\usepackage{microtype}

\usepackage{import}
\usepackage{graphicx}
\usepackage{wasysym}
\usepackage{dcolumn}
\usepackage{bm}
\usepackage{booktabs}
\usepackage{siunitx}
\usepackage{xcolor}
\usepackage{comment}

\usepackage[normalem]{ulem}
\usepackage[version=4]{mhchem} 
\usepackage{lipsum} 
\usepackage{textcomp} 

\usepackage{url}
\usepackage{xurl}

\definecolor{crossrefBlue}{RGB}{0, 84, 159}
\definecolor{citeBlue}{RGB}{100, 100, 100}

\usepackage[
    colorlinks=true,
    linkcolor=crossrefBlue,  
    citecolor=citeBlue,      
    urlcolor=crossrefBlue   
]{hyperref}

\hypersetup{breaklinks=true}
\usepackage{cleveref}
\crefname{figure}{FIG.}{FIGS.}
\crefname{table}{TABLE}{TABLES}
\crefname{equation}{Eq.}{Eqs.}

\begin{document}

\newcolumntype{C}[1]{>{\centering\let\newline\\\arraybackslash\hspace{0pt}}m{#1}}

\title{
    Phononic Bragg reflectors for thermal insulation\\
    of scalable cryogenic control electronics from qubits
}

\author{Isabelle V. Sprave}
\author{Denny Dütz}
\author{Sebastian Kock}
\author{René Otten}
\author{Tobias Hangleiter}
\affiliation{JARA-FIT Institute for Quantum Information, Forschungszentrum J\"ulich GmbH and RWTH Aachen University, 52074~Aachen, Germany}
\author{Felix Mende}
\affiliation{Fraunhofer-Institut f\"ur Photonische Mikrosysteme IPMS, 01109~Dresden, Germany}
\affiliation{Institute of Applied Physics, TU Dresden, Nothnitzer Strasse 61, 01187~Dresden, Germany}
\author{Marcus Wislicenus}
\affiliation{Fraunhofer-Institut f\"ur Photonische Mikrosysteme IPMS, 01109~Dresden, Germany}
\author{Hendrik Bluhm}
\email{bluhm@physik.rwth-aachen.de}
\affiliation{JARA-FIT Institute for Quantum Information, Forschungszentrum J\"ulich GmbH and RWTH Aachen University, 52074~Aachen, Germany}
\affiliation{ARQUE Systems GmbH, 52074~Aachen, Germany}

\date{\today}

\definecolor{Red}{rgb}{1,0,0}
\newcommand{\add}[1]{#1}
\newcommand{\del}[1]{\textcolor{Red}{\sout{}}}

\begin{abstract}
Scaling solid-state architectures to the millions of qubits required for utility-scale quantum computing could benefit from the integration of control electronics in the immediate vicinity of the quantum layer. 
However, lithographically fabricated solid-state qubits perform best at temperatures well below $\SI{1}{K}$, where available cooling power is limited, whereas the control electronics dissipate substantial power and therefore require the higher cooling power available at elevated temperatures. 
To address this challenge, we propose a cryopackaging concept that uses broadband phononic Distributed Bragg Reflectors (DBRs) as a thermal barrier between cryoelectronics and the qubit chip. As an experimental realization of this concept, we fabricate and characterize Ta/\ce{SiO2} DBR structures. 
In this architecture, the DBR is intended to provide mechanical support for superconducting vias while offering substantially better thermal insulation than typical bulk materials. 
For a 600-nm-thick DBR consisting of 10~Ta/\ce{SiO2} bilayers, we obtain a thermal conduction below $\SI{1}{mW/cm^2}$ from $\SI{1.5}{K}$ to $\SI{100}{mK}$. 
In a centimeter-scale architecture, this level of isolation is compatible with Watt-level cooling power for nearby electronics while maintaining a qubit temperature around $\SI{100}{mK}$ in commercially available dilution refrigerators. 
\end{abstract}

\maketitle

\section{Introduction} \label{sec:Intro}

Solid-state qubits such as superconducting and semiconductor qubits are controlled by electrical signals in the microwave or sub-microwave domain. These types of qubits  typically require several control signals per qubit and benefit from operation at sub-Kelvin temperatures. The number of required control signals ranges from 2 (for NV-centers in diamonds) to up to about 10 lines per qubit (for shuttling-based quantum dot architectures) \cite{NV_center, Transmon, Donor_Spin_Qubit, Spiderweb, SpinBus_Architecture}.
In current experiments, these signals are usually generated by discretely wired room-temperature electronics.
Extending this approach to the millions of qubits expected to be needed for practical quantum computing applications would thus lead to an impractically large number of cables.

%------------- Fig 1 Vision ---------------
\begin{figure}[t!] 
    \centering
    \def\svgwidth{4in}
    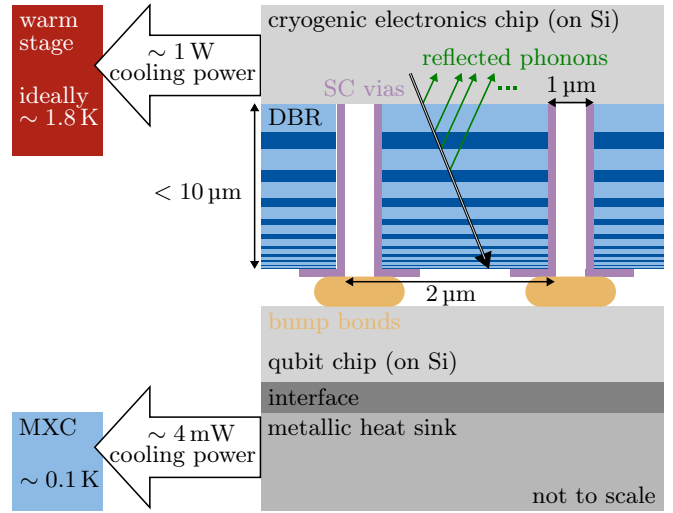
    \caption{\label{fig:Fig1_Vision}
    \textbf{Concept of cryopackaging for close-to-qubit control (not to scale).}
    Schematic of the proposed architecture, in which the cryoelectronics chip is thermally anchored to a warmer stage with higher cooling power, while the qubit chip remains connected to the mixing chamber (MXC) stage at $\SI{100}{mK}$. A phononic distributed Bragg reflector (DBR) stack acts as a thermal barrier between the chips while accommodating dielectric lining and superconducting (SC) through-vias for dense electrical interconnects. The DBR reflects phonons emitted by the cryoelectronics, thereby reducing heat flow toward the qubit layer. All numbers shown are design targets or estimates rather than measured values (see also Appendix \ref{sec:Estimates_concept}).
    }
\end{figure}
%-------- ENDE Fig 1 Vision ----------------

An attractive solution to avoid this wiring complexity is to implement part of the control functionality in the immediate vicinity of the qubits \cite{Xue,Cryo-CMOS} such that a high density of interconnects can be realized via microfabrication techniques. 
The functionality of the local control electronics must then be sufficient to  reduce the number of macroscopic signal lines connected to room-temperature electronics to a manageable level.  
A major challenge for such an integration is the limited cooling power of cryostats, which decreases with mixing chamber temperature and thus limits the power budget of cryogenic electronics in order to not elevate the qubit operating temperature.
Semiconductor spin qubits are typically operated at a few hundred millikelvin \cite{mills_high-fidelity_2022, 200mK_hotter_is_easier}, and, although operation above $\SI{1}{K}$ has been demonstrated, performance degrades at higher temperatures \cite{high-fidelity_performance_better_below_1K}. Superconducting qubits even require temperatures well below $\SI{100}{mK}$ for a good performance \cite{20mK_superconducting_qubits}. 

We propose to address these conflicting requirements by introducing a thermal insulation matrix between qubits and electronics that still allows the realization of high density microfabricated interconnects (see \Cref{fig:Fig1_Vision}).
To realize a matrix satisfying the requirements stated below on thermal insulation, mechanical support and via compatibility, we consider a Distributed Bragg Reflector (DBR) that reflects phonons very efficiently over the full relevant thermal frequency spectrum. 
We demonstrate that at $\SI{1.8}{K}$, 600-nm- and 440-nm-thick Ta/\ce{SiO2} DBRs with 10 bilayers each reduce the heat flow toward $\SI{100}{mK}$ to below $\SI{2}{\milli W \per cm^2}$ and $\SI{3}{\milli W \per cm^2}$, respectively. At $\SI{1.5}{K}$, both DBR designs reduce the heat flow to below $\SI{1}{\milli W \per cm^2}$. 

Using such an insulation, the cryoelectronics can be cooled by a warmer refrigerator stage with greater cooling power, 
for example by exploiting the different temperature stages of a dilution refrigerator. One could also introduce a dedicated warm cooling stage around, e.g., $\SI{1.8}{K}$, where cooling powers in the order of $\SI{100}{W}$ can be achieved in commercially available refrigerators or even kilowatt in large scale superfluid helium systems \cite{1.8K_superfluid_He}. 
However, note that, in practice, one is limited by the intrinsic thermal conduction of Si in our use case, so that a Watt-level cooling power is more realistic. 

The thermal conduction of the interconnects to the qubits must then be low enough for the heat flow not to exceed the cooling power available at the target temperature. In commercial dilution refrigerators this cooling power amounts to a few milliwatts at $\SI{100}{mK}$ (e.g. Leiden Cryogenics CF-CS-XXL\cite{Leiden_cooling_power}). 

The three-dimensional (3D) integration scheme that we propose consists not only of a thin, thermally insulating matrix but also of high-density superconducting vias hosted by the DBR that is positioned between the cryogenic electronics and the quantum layer. While vacuum provides excellent thermal insulation, 
a solid matrix is required for mechanical stability of the vias.
Unlike lateral integration \cite{pauka_cryogenic_2021_lateral_integration, bartee_spin-qubit_2025}, the proposed 3D configuration supports a high interconnect density at low levels of the system stack, enabling scalability to millions of qubits. 
A similar route to thermal isolation via heterogeneous packaging employs flexible interposers \cite{bickel_flex_interposer}. Thanks to their long, potentially superconducting  signal lines microfabricated on a thin organic support, they should achieve a similar level of thermal isolation, but are ultimately also limiting the wiring density.

A DBR is made up of alternating layers of materials with contrasting physical properties, designed to achieve high reflectivity within a certain target spectral range by leveraging interference effects. The concept originates from optics, where alternating layers with different refractive indices form \textit{photonic} DBRs that reflect light. 
The DBR's performance depends on the layer design: Constant layer thicknesses yield narrow reflection bands centered on a target wavelength (periodic design, e.g., \cite{optical_DBR, optical_DBR_GaAs}), while varying layer thicknesses broaden the reflection range through constructive interference across multiple wavelengths (aperiodic, graded, or chirped designs, e.g., \cite{optical_DBR_double-chirped}).
The same principle applies to acoustic systems, where \textit{phononic} DBRs reflect mechanical waves through discontinuities not in refractive index, but in acoustic impedance $Z = \varrho \cdot v$, which depends on the material density $\varrho$ and speed of sound $v$. Acoustic DBRs are widely used in radio frequency (RF) filters \cite{acoustic_DBR}. 
Superconducting acoustic Bragg reflectors have been reported in Ref.~\cite{bon-mardion_superconducting_2024}, combining niobium (Nb) and titanium nitride (TiN) in a periodic configuration to suppress phononic heat transport with a similar motivation as ours.
Dütz et al. \cite{Denny_DBR} conducted simulations of DBRs composed of silicon dioxide (\ce{SiO2}) combined with tantalum (\ce{Ta}), tungsten (\ce{W}), or iridium (\ce{Ir}). They  concluded that varying the layer thicknesses significantly enhances broadband constructive interference, yielding much stronger phonon reflection compared to periodic designs.

This paper aims to experimentally verify the predictions made by Dütz \cite{Denny_DBR} (see also \Cref{sec:Theory_DBR}).
Our complete cryopackaging concept for close-to-qubit control is illustrated in \Cref{fig:Fig1_Vision}. The target performance and geometric parameters are derived from estimates for a concrete implementation scenario based on the SpinBus architecture (Ref.~\cite{SpinBus_Architecture}); the underlying assumptions are discussed in Appendix \ref{sec:Estimates_concept}.

This paper is organized as follows:
\Cref{sec:Theory_DBR} briefly reviews the main results of Ref.~\cite{Denny_DBR} in the interest of self-containment of this work. 
The sample preparation and experimental methods are described in \Cref{sec:Material_Methods}. Subsequently, we present the experimental data and analysis in \Cref{sec:Analysis}, followed by the results in \Cref{sec:Results} and discussion in \Cref{sec:discussion_conclusion}. 
Additional details on the simulation and growth of DBRs, the setup and measurement procedure used, as well as the discussion of potential error sources are provided in the Appendix.

\section{Predicted performance of broadband phononic Bragg Reflectors}
\label{sec:Theory_DBR}

%------------- Fig 2 Sample and Setup ---------------
\begin{figure*}[t!]
    \centering
    \def\svgwidth{7in}
    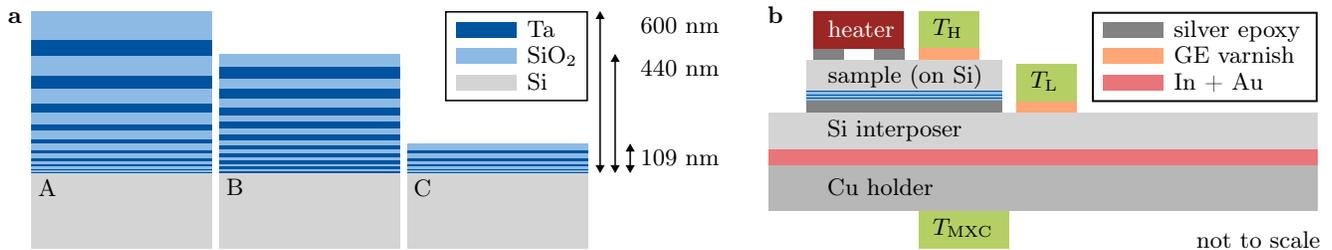
    \caption{\label{fig:Fig2_Setup_Samples}
    \textbf{Experimental Setup. }
    \textbf{a} Overview of the three DBR sample types investigated: the total Ta/\ce{SiO2} thicknesses are $\SI{600}{nm}$ (A), $\SI{440}{nm}$ (B), and $\SI{109}{nm}$ (C). In addition, three reference samples consisting of bare Si (D) and \ce{SiO2} layers of $\SI{503}{nm}$ (E) and $\SI{1438}{nm}$ (F) on Si substrates are measured (samples D, E, and F are described in \Cref{tab:Samples}). Sample E serves as the bulk material reference for sample B due to its comparable thickness, while sample F is more than twice as thick as the thickest DBR (A). At least two samples were measured for each type.
    \textbf{b} Schematic of the experimental configuration. Each sample is mounted with the DBR or reference oxide facing away from the heater, on an Si interposer with a $\SI{200}{\nano m}$-thin gold-coated backside. Both Si pieces are intended to homogenize the temperature laterally, so that a uniform temperature is maintained even if the heat is applied locally. The interposer is attached to a gold-plated oxygen-free high-conductivity (OFHC) copper holder that is tightly screwed onto the mixing chamber (MXC) plate of the dilution refrigerator. A $\SI{50}{\micro m}$ indium foil between the gold surfaces provides enhanced Cu–Si thermal contact. The sample is bonded to the interposer using silver epoxy, and a resistive heater is mounted on the top surface. During measurements, the heater power is varied between $\SI{0}{W}$ and $\SI{1}{mW}$, and temperatures are monitored on the sample, interposer, and holder ($T_{\mathrm{H}}$, $T_{\mathrm{L}}$, and $T_\mathrm{MXC}$). The MXC temperature is stabilized via a PID loop at $\SI{100}{mK}$ to $\SI{700}{mK}$ in steps of $\SI{100}{mK}$.
     }
\end{figure*}
%-------- ENDE Fig 2 Sample and Setup ----------------

The theoretical foundation of this work is provided in Ref.~\cite{Denny_DBR}, which presents a detailed numerical model for phononic DBRs. In the following, we summarize the essential concepts and main results that guided the design investigated in this paper.

To achieve broadband acoustic reflection, the DBRs we studied employ alternating layers of materials with large acoustic impedance contrast, while the individual layer thicknesses are gradually varied. This graded design enhances reflection over a wide frequency range, but it also increases the overall number of layers and the total stack thickness, requiring a compromise between reflectivity and compactness.

To determine an optimal configuration, the phonon transmission for all relevant angles and frequencies is computed and the results are integrated over the thermal phonon spectrum. The thermal transfer was then evaluated as a function of geometry, revealing that exponentially increasing layer thicknesses yield the best broadband reflection performance out of the geometries tested. This variable thickness profile is in contrast with the superconducting superlattice presented in Ref.~\cite{bon-mardion_superconducting_2024}, which has constant layer thicknesses (periodic geometry). 

Several material combinations were investigated, including \ce{SiO2} paired with tantalum, tungsten, or iridium (see also \Cref{tab:Impedances_Simulation} in Appendix~\ref{sec:Simulations_LogLog}). Assuming that coherent phonons dominate thermal transport, the simulations predict a suppression of the transmitted power as $T \rightarrow \SI{0}{K}$. 

Among the studied material combinations, Ir/\ce{SiO2} and W/\ce{SiO2} DBRs yield the lowest simulated power transmission at $\SI{1.4}{K}$, reaching $\SI{0.3}{\micro W/cm^2}$ and $\SI{0.9}{\micro W/cm^2}$, respectively, for a total thickness of $\SI{7}{\micro m}$. 
For the experimental investigation, however, we focus on Ta/\ce{SiO2} DBRs, as tantalum is widely used in semiconductor fabrication and thin-film growth is readily available at Fraunhofer IPMS (see Appendix~\ref{sec:growth}). 
The simulated power transmission for this material system remains sufficiently low, about $\SI{0.1}{\milli W/cm^2}$ for a thickness of $\SI{7}{\micro m}$, and thus still satisfies the thermal requirements outlined above. 
Importantly, the predicted heat transfer through DBRs with total thicknesses below $\SI{10}{\micro m}$ remains well below the target value of $\SI{1}{mW/cm^2}$ at $\SI{1.8}{K}$. This transmitted power is more than two orders of magnitude smaller than that through $\SI{10}{\micro m}$-thick layers of the best available bulk materials (graphite or Vespel SP-22), which exceed $\SI{100}{mW/cm^2}$ at $\SI{1.8}{K}$ \cite{LakeShoreAppendixI,Vespel}. In the following, we experimentally investigate Ta/\ce{SiO2} DBR structures to verify these predictions and assess their suitability for the proposed cryopackaging concept.

\section{Experimental implementation} \label{sec:Material_Methods}

\subsection{Samples}
\label{subsec:Samples}

\renewcommand{\arraystretch}{1.2}
\begin{table}[t]
    \centering
    \caption{\textbf{Overview of the six sample types investigated.} All samples share the same ca.~$\SI{750}{\micro m}$ bare Si (100) substrate with room temperature electrical resistivity  values of approximately $\SI{15}{\ohm\ \centi\metre}$. The total thickness indicates the added material layers. The number of measured samples is equal to the number of dark blue / light blue lines in \Cref{fig:Fig4_Result}.} 
    \label{tab:Samples}
    \begin{ruledtabular}
        \begin{tabular}{ccccc}
           \shortstack{sample\\type}  &\raisebox{0.3\normalbaselineskip}{material(s)} & \raisebox{0.3\normalbaselineskip}{\# bilayers} & \shortstack{total thickness \\ in nm} & \shortstack{\# meas. \\ samples}\\
           \hline
           A   & Ta/\ce{SiO2} & 10 & 600 & 3\\
           B   & Ta/\ce{SiO2} & 10 & 440 & 2\\
           C   & Ta/\ce{SiO2} & 5 & 109 & 2\\
           D   & -            & -  &  -   & 4\\
           E   & \ce{SiO2}   & -  & 503 & 3\\
           F    & \ce{SiO2}  & -  & 1438 & 3\\
        \end{tabular}
    \end{ruledtabular}
\end{table}

Figure~\ref{fig:Fig2_Setup_Samples}\textcolor{crossrefBlue}{\textbf{a}} illustrates the three Ta/\ce{SiO2} DBRs investigated in this study.  All structures were deposited on the same $\SI{750}{\micro m}$-thick Si substrate (100-direction parallel to surface normal) 
and have lateral dimensions of approximately $\SI{3.5}{mm} \times \SI{3.5}{mm}$.

An overview of the sample types as well as three reference samples and their total thicknesses is given in \Cref{tab:Samples}. Further details on the deposition process and layer growth parameters are provided in Appendix \ref{sec:growth}.

\subsection{Method overview}

To directly assess the thermal insulating effect of the DBRs, we establish a temperature gradient across the sample, with a higher temperature $T_{\mathrm{H}}$ on the top side and a lower temperature $T_{\mathrm{L}}$ at the bottom (see \Cref{fig:Fig2_Setup_Samples}\textcolor{crossrefBlue}{\textbf{b}} and \Cref{sec:Measurement_Procedure}). This gradient is generated by applying a heater power $P$ to the top surface, while the bottom surface is thermally anchored to the mixing chamber (MXC) of a dilution refrigerator.
We perform the measurements at seven different MXC temperatures $T_{\mathrm{MXC}} \in \{\SI{100}{mK}, \SI{200}{mK}, ..., \SI{700}{mK}\}$ to later check for consistency (see \Cref{fig:Fig3_Verification}).
After applying the heating power, the system is allowed to reach a steady state. Subsequently, the temperatures $T_{\mathrm{H}}$, $T_{\mathrm{L}}$, and $T_{\mathrm{MXC}}$ are recorded. Because of additional thermal impedances within the setup, $T_{\mathrm{L}}$ varies along with $T_{\mathrm{H}}$ even when $T_{\mathrm{MXC}}$ remains constant (see \Cref{fig:FigB_Rawdata_2}). Additionally, $T_{\mathrm{MXC}}$ increases once $P$ exceeds the available cooling power of the dilution refrigerator. These variations are accounted for in the determination of the heat flow toward the MXC and the effective thermal conductivity, as described in \Cref{sec:Analysis}.

Details on the setup and the measurement procedure are presented in Appendix \ref{subsec:Setup}, \ref{sec:Si-Cu-interface}, and \ref{sec:Measurement_Procedure}, respectively.
Potential error sources are discussed in Appendix \ref{sec:ThermometryMethods}.

% =======================================================================================

\section{Experimental Data \& analysis} \label{sec:Analysis}

\begin{figure}
    \centering
    \includegraphics[width=1\linewidth]{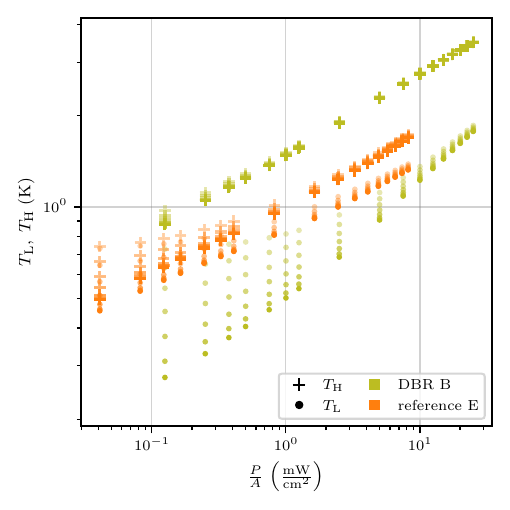}
    \caption{\label{fig:FigB_Rawdata_2}
    \textbf{Exemplary raw data.} 
    Calibrated raw data for two samples of comparable thickness: DBR B (green) and reference E (orange, see \Cref{tab:Samples} and \Cref{fig:Fig2_Setup_Samples}\textcolor{crossrefBlue}{\textbf{a}}. Even though the applied heating power $P$ was increased from $\SI{0}{W}$ to $\SI{1}{mW}$ in both cases, the values of $P/A$ vary due to slight differences in sample dimensions. The data points for $P~=~\SI{0}{W}$ are not visible because of the logarithmic axes.
    The different transparencies correspond to fixed mixing chamber temperatures $T_{\mathrm{MXC}}$ between $\SI{100}{mK}$ (solid color) and $\SI{700}{mK}$ (most transparent color). 
    The data show that both, $T_{\mathrm{H}}$ (+) and $T_{\mathrm{L}}$ ($\bullet$), vary with increasing $P$, indicating that a self-consistent analysis is required to extract the thermal sample properties according to \Cref{eq:heat_flow}.
    }
\end{figure}

Figure~\ref{fig:FigB_Rawdata_2} shows exemplary measured temperatures $T_{\mathrm{L}}$ and $T_{\mathrm{H}}$ at the cold and hot sides of the samples, respectively. 
To quantify the thermal insulation effect of the DBRs, these quantities measured at a fixed $T_{\mathrm{MXC}}$, as well as the applied heating power $P$, must be converted into a geometry-independent quantity that reflects the ability of the interface to transmit heat as a function of temperature.
The temperature-dependent thermal conductivity $\lambda(T)$ is in principle such a quantity. However, $\lambda(T)$ is not well-defined inside the DBR as the phonon distribution is not expected to be thermal so that the temperature in the DBR is ill-defined.

According to the quasi-ballistic heat transfer model used in Ref.~\cite{Denny_DBR}, which assumes that phonons only get reflected elastically and coherently while traversing the DBR, 
the net heat flux from the warm to the cold side of a sample of cross-sectional area $A$ and thickness $L$ is given by the difference between the heat integrals $I(T)$ in both directions, each of which only depend on the source temperature:
\begin{equation} \label{eq:heat_flow}
    \frac{P}{A} = \frac{I(T_{\mathrm{H}}(P, T_{\mathrm{MXC}}))}{L} - \frac{I(T_{\mathrm{L}}(P, T_{\mathrm{MXC}}))}{L} \medspace .
\end{equation}
This form is a direct consequence of neglecting phonon-phonon and phonon-electron interaction. 
\Cref{eq:heat_flow} is of the same form as the integral form of Fourier’s law, which applies in the diffusive regime: 
 \begin{equation} \label{eq:diffusive}
     \frac{P}{A} = \frac{1}{L} \cdot \int_{T_{\mathrm{L}}}^{T_{\mathrm{H}}} \lambda(T) dT  \medspace \rightarrow I(T) = \int_0^T \lambda(T') dT' + C\medspace .
\end{equation}

It thus allows to define the \textit{effective} thermal conductivity of the sample $\lambda_{\mathrm{eff}}(T)$ as the  derivative of $I(T)$. $\lambda_{\mathrm{eff}}(T)$ can be used for comparison with bulk materials. Note that the constant $C$ has neither an influence on  $\lambda_{\mathrm{eff}}(T)$, nor on $P/A$ as $C$ cancels out in the difference in \Cref{eq:heat_flow}. 

\begin{figure}
    \centering
    \def\svgwidth{3.4in}
    %% Creator: Inkscape 1.2.2 (b0a8486, 2022-12-01), www.inkscape.org
%% Accompanies image file 'Fig4_Analysis_abc.pdf' (pdf, eps, ps)
\begingroup%
  \makeatletter%
  \providecommand\color[2][]{%
    \errmessage{(Inkscape) Color is used for the text in Inkscape, but the package 'color.sty' is not loaded}%
    \renewcommand\color[2][]{}%
  }%
  \providecommand\transparent[1]{%
    \errmessage{(Inkscape) Transparency is used (non-zero) for the text in Inkscape, but the package 'transparent.sty' is not loaded}%
    \renewcommand\transparent[1]{}%
  }%
  \providecommand\rotatebox[2]{#2}%
  \newcommand*\fsize{\dimexpr\f@size pt\relax}%
  \newcommand*\lineheight[1]{\fontsize{\fsize}{#1\fsize}\selectfont}%
  \ifx\svgwidth\undefined%
    \setlength{\unitlength}{244.79999734bp}%
    \ifx\svgscale\undefined%
      \relax%
    \else%
      \setlength{\unitlength}{\unitlength * \real{\svgscale}}%
    \fi%
  \else%
    \setlength{\unitlength}{\svgwidth}%
  \fi%
  \global\let\svgwidth\undefined%
  \global\let\svgscale\undefined%
  \makeatother%
  \begin{picture}(1,1.77380032)%
    \lineheight{1}%
    \setlength\tabcolsep{0pt}%
    \put(0,0){\includegraphics[width=\unitlength,page=1]{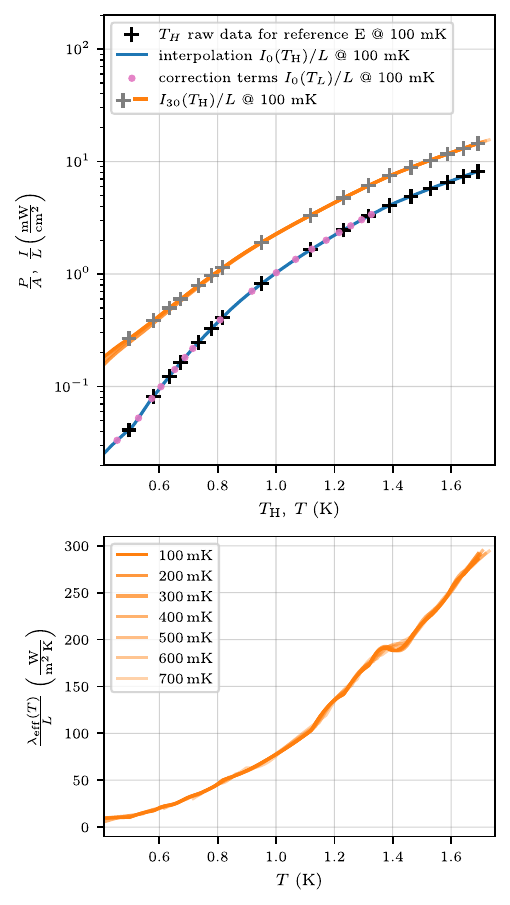}}%
    \put(0.02205593,1.72634682){\color[rgb]{0,0,0}\makebox(0,0)[lt]{\lineheight{1.25}\smash{\begin{tabular}[t]{l}\textbf{a}\end{tabular}}}}%
    \put(0.02164753,0.6954506){\color[rgb]{0,0,0}\makebox(0,0)[lt]{\lineheight{1.25}\smash{\begin{tabular}[t]{l}\textbf{b}\end{tabular}}}}%
  \end{picture}%
\endgroup%

    \caption{\label{fig:Fig3_Analysis}
    \textbf{Iterative analysis of experimental data for sample E at $\SI{100}{mK}$ based on \Cref{eq:heat_flow}.}
    \textbf{a}~Illustration of the iterative correction procedure. Black crosses show the measured data, the blue line the initial PCHIP interpolation, pink dots the interpolated correction terms, and grey crosses with orange line the corrected data and final interpolation after 30 iterations. Measurements at seven mixing chamber temperatures between $\SI{100}{mK}$ and $\SI{700}{mK}$ demonstrate consistent results; curves for $T_{\mathrm{MXC}} \geq \SI{200}{mK}$ are shifted by constant offsets as described in the text.
    \textbf{b}~Effective thermal conductivity $\lambda_{\mathrm{eff}}(T)$ obtained from the derivative of the final $I(T)$. The agreement between curves derived from different $T_{\mathrm{MXC}}$ confirms the consistency of the method.
    }
\end{figure}

As shown in \Cref{fig:FigB_Rawdata_2}, both $T_{\mathrm{H}}$ and $T_{\mathrm{L}}$ vary during each measurement sequence. Because the heat flux $P/A$ depends on both temperatures, $I(T)$ cannot be obtained directly and must instead be determined self-consistently, which we achieve through an iterative procedure.
Figure~\ref{fig:Fig3_Analysis}\textcolor{crossrefBlue}{\textbf{a}} illustrates the first step and last result of this iterative analysis method.

%------------- Fig 3 Raw Data ---------------
%-------- ENDE Fig 3 Raw Data ----------------

%------------- Fig 4 Analysis ---------------
%--------ENDE Fig 4 Analysis----------------

%------------- Fig 5 Verification ---------------
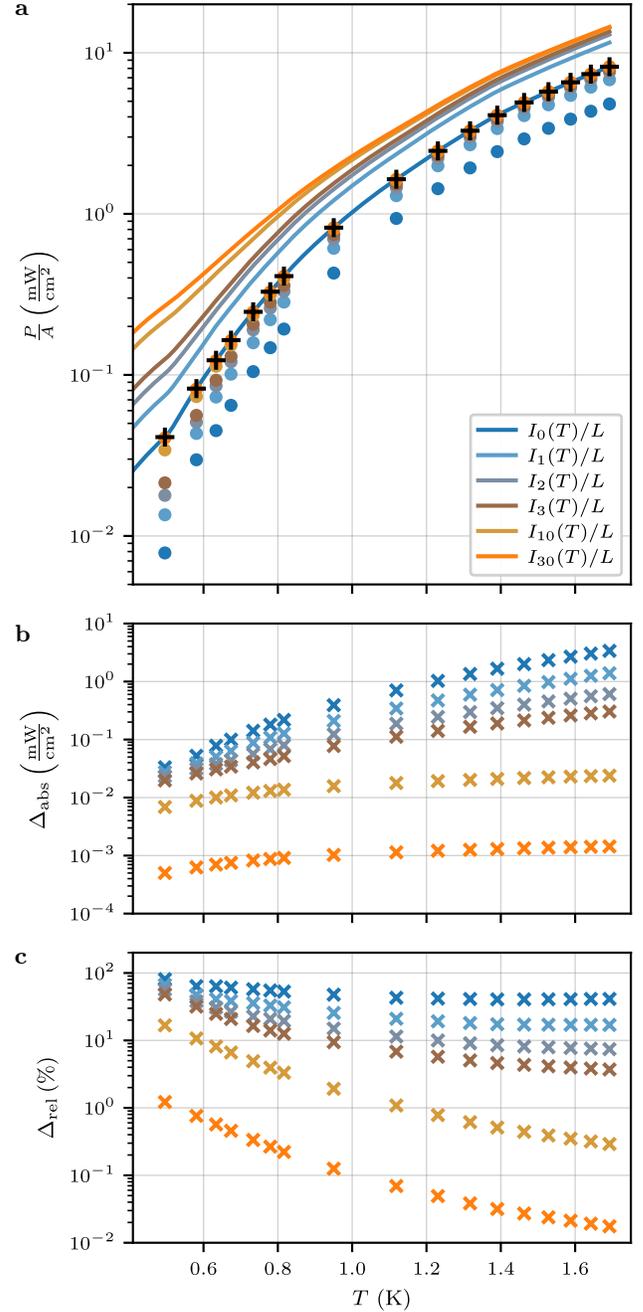
\begin{figure}
    \centering
    \def\svgwidth{3.4in}
    %% Creator: Inkscape 1.2.2 (b0a8486, 2022-12-01), www.inkscape.org
%% Accompanies image file 'Fig5_Verification_abc.pdf' (pdf, eps, ps)
\begingroup%
  \makeatletter%
  \providecommand\color[2][]{%
    \errmessage{(Inkscape) Color is used for the text in Inkscape, but the package 'color.sty' is not loaded}%
    \renewcommand\color[2][]{}%
  }%
  \providecommand\transparent[1]{%
    \errmessage{(Inkscape) Transparency is used (non-zero) for the text in Inkscape, but the package 'transparent.sty' is not loaded}%
    \renewcommand\transparent[1]{}%
  }%
  \providecommand\rotatebox[2]{#2}%
  \newcommand*\fsize{\dimexpr\f@size pt\relax}%
  \newcommand*\lineheight[1]{\fontsize{\fsize}{#1\fsize}\selectfont}%
  \ifx\svgwidth\undefined%
    \setlength{\unitlength}{244.79999734bp}%
    \ifx\svgscale\undefined%
      \relax%
    \else%
      \setlength{\unitlength}{\unitlength * \real{\svgscale}}%
    \fi%
  \else%
    \setlength{\unitlength}{\svgwidth}%
  \fi%
  \global\let\svgwidth\undefined%
  \global\let\svgscale\undefined%
  \makeatother%
  \begin{picture}(1,2.07141707)%
    \lineheight{1}%
    \setlength\tabcolsep{0pt}%
    \put(0,0){\includegraphics[width=\unitlength,page=1]{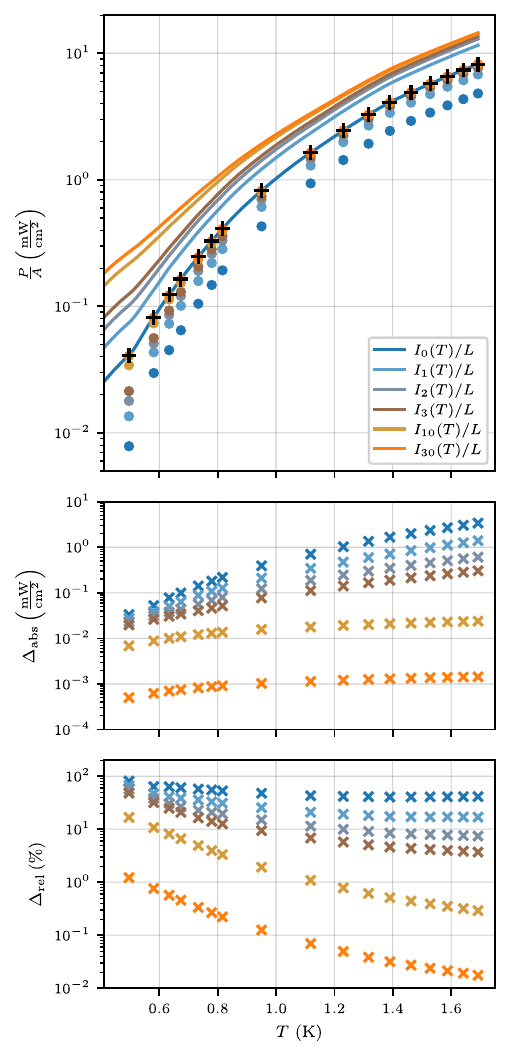}}%
    \put(0.02205593,2.02454254){\color[rgb]{0,0,0}\makebox(0,0)[lt]{\lineheight{1.25}\smash{\begin{tabular}[t]{l}\textbf{a}\end{tabular}}}}%
    \put(0.02205593,1.05997331){\color[rgb]{0,0,0}\makebox(0,0)[lt]{\lineheight{1.25}\smash{\begin{tabular}[t]{l}\textbf{b}\end{tabular}}}}%
    \put(0.02160658,0.56320625){\color[rgb]{0,0,0}\makebox(0,0)[lt]{\lineheight{1.25}\smash{\begin{tabular}[t]{l}\textbf{c}\end{tabular}}}}%
  \end{picture}%
\endgroup%

    \caption{\label{fig:Fig3_Verification}
    \textbf{Verification of the iterative analysis method for sample E at $\SI{100}{mK}$.}
    \textbf{a}~Heat integrals $I_n(T)$ obtained after successive iterations (continuous lines). In Addition, calculated values $[I_n(T_{\mathrm{H}})-I_n(T_{\mathrm{L}})]$ ($\bullet$) are compared with the measured data (+), showing good agreement for $n=30$.
    \textbf{b}~Absolute deviation between the measured values and the recalculated heat flow decreases with iteration number.
    \textbf{c}~Relative deviation between measurement and model. After 30 iterations (orange), the deviation falls below $\SI{2}{\percent}$, indicating convergence of the method.
    }
\end{figure}
%--------ENDE Fig 5 Verification----------------

In each iteration $n~\in~\mathbb{N}^+$, the correction term $I(T_{\mathrm{L}})$ is replaced by the estimate from the previous iteration:
\begin{equation} \label{eq:I_n}
    \frac{I_n(T_{\mathrm{H}})}{L} = \frac{P}{A} + \frac{I_{n-1}(T_{\mathrm{L}})}{L} \quad .
\end{equation}

For the initial iteration, $I_0(T_{\mathrm{L}})$ is set to zero:
\begin{equation}
    \frac{P}{A} \approx \frac{I(T_{\mathrm{H}})}{L} = \frac{I_0(T_{\mathrm{H}})}{L} \medspace .
\end{equation}
This starting value already gives a reasonable approximation because the thermal conductivity $\lambda(T)$ increases steeply at sub-Kelvin temperatures, leading to $I(T_{\mathrm{H}}) \gg I(T_{\mathrm{L}})$.
When the sequence defined by \Cref{eq:I_n} converges such that $I_n(T_{\mathrm{H}}) \approx I_{n-1}(T_{\mathrm{H}})$, the desired function $I(T)$ is obtained. 

For the iterative approach to work, some form of interpolation is needed so that the correction terms can be evaluated at arbitrary temperatures.
According to the behavior predicted by simulations (see \Cref{fig:FigA_Simulations_LogLog} in Appendix \ref{sec:Simulations_LogLog}), a simple power-law fit across the full temperature range is inadequate: in log–log space, the data clearly deviates from a single linear dependence and instead exhibits three distinct linear regimes. 
Therefore, we employ a Piecewise Cubic Hermite Interpolating Polynomial (PCHIP). This type of interpolation allows us to interpolate the datapoints in a continuously differentiable manner, without unphysical overshoots. 

To verify the convergence of the iterative analysis, we compare the experimentally measured heat flux $P/A$ with the values recalculated from the extracted heat integral $I_n(T)$ after each iteration (Fig.~\ref{fig:Fig3_Verification}). For iteration $n$, the predicted heat flow is obtained as $I_n(T_{\mathrm{H}})-I_n(T_{\mathrm{L}})$ and compared with the measured power density $P/A$. We quantify the agreement using the absolute deviation
\begin{equation}
    \Delta_{\mathrm{abs}} = \mathrm{raw\,data} - [I_n(T_{\mathrm{H}}) - I_n(T_{\mathrm{L}})]
\end{equation}
and the relative deviation
\begin{equation}
    \Delta_{\mathrm{rel}} = \frac{\Delta_{\mathrm{abs}}}{\mathrm{raw\,data}} .
\end{equation}
As shown in Fig.~\ref{fig:Fig3_Verification}\textcolor{crossrefBlue}{\textbf{b}}, the absolute deviation decreases monotonically with iteration number. After 30 iterations, the relative deviation falls below $\SI{2}{\percent}$ for all data points (Fig.~\ref{fig:Fig3_Verification}\textcolor{crossrefBlue}{\textbf{c}}), confirming that \Cref{eq:heat_flow} can reproduce the experimental data and that the procedure has converged. We therefore use this threshold as the stopping criterion for the iterative procedure. 

To  further verify the robustness of the approach, the measurements were repeated at seven mixing chamber temperatures between $\SI{100}{mK}$ and $\SI{700}{mK}$, controlled via a PID loop. For each $T_{\mathrm{MXC}}$, the iterative procedure yields a curve $I_n^{T_{\mathrm{MXC}}}(T_{\mathrm{H}})$, which converges to the integrated thermal conductance $I^{T_{\mathrm{MXC}}}(T)$. Since the experimentally accessible temperature range starts at the respective base temperature $T_{\mathrm{MXC}}$, the curves obtained for $T_{\mathrm{MXC}} \geq \SI{200}{mK}$ differ from the $\SI{100}{mK}$ data by an additive constant offset  $C^{T_{\mathrm{MXC}}}$. To compare them on the same scale, we shift these curves by $C^{T_{\mathrm{MXC}}}$ according to \Cref{eq:diffusive}, choosing
\begin{equation}
    C^{T_{\mathrm{MXC}}} = I_{30}^{\SI{100}{mK}}\!\left(T_{\mathrm{MXC}}(P=\SI{0}{W})\right).
\end{equation}
With this choice, all curves agree well over the common temperature range, confirming the consistency of the analysis (see \Cref{fig:Fig3_Analysis}\textcolor{crossrefBlue}{\textbf{a}}). The remaining differences at the lowest temperatures reflect the arbitrary choice of integration constant due to the limited low-temperature access of each measurement, and do not affect the extracted thermal transport properties. 

Finally, referencing the integral form of Fourier's law, the \textit{effective} thermal conductivity $\lambda_{\mathrm{eff}}(T)$ is derived analytically from the heat integral $I(T)$: 
\begin{equation}
    \lambda_{\mathrm{eff}}(T) = \frac{d I(T)}{dT} \medspace .
\end{equation}

The resulting temperature dependence of $\lambda_{\mathrm{eff}}(T)$ is presented in \Cref{fig:Fig3_Analysis}\textcolor{crossrefBlue}{\textbf{b}}.

% =======================================================================================

\section{Results \& Discussion} \label{sec:Results}

\begin{figure}[t!]
    \centering
    \def\svgwidth{3.4in}
    %% Creator: Inkscape 1.2.2 (b0a8486, 2022-12-01), www.inkscape.org
%% Accompanies image file 'Fig6_Result_abc.pdf' (pdf, eps, ps)
\begingroup%
  \makeatletter%
  \providecommand\color[2][]{%
    \errmessage{(Inkscape) Color is used for the text in Inkscape, but the package 'color.sty' is not loaded}%
    \renewcommand\color[2][]{}%
  }%
  \providecommand\transparent[1]{%
    \errmessage{(Inkscape) Transparency is used (non-zero) for the text in Inkscape, but the package 'transparent.sty' is not loaded}%
    \renewcommand\transparent[1]{}%
  }%
  \providecommand\rotatebox[2]{#2}%
  \newcommand*\fsize{\dimexpr\f@size pt\relax}%
  \newcommand*\lineheight[1]{\fontsize{\fsize}{#1\fsize}\selectfont}%
  \ifx\svgwidth\undefined%
    \setlength{\unitlength}{244.79999734bp}%
    \ifx\svgscale\undefined%
      \relax%
    \else%
      \setlength{\unitlength}{\unitlength * \real{\svgscale}}%
    \fi%
  \else%
    \setlength{\unitlength}{\svgwidth}%
  \fi%
  \global\let\svgwidth\undefined%
  \global\let\svgscale\undefined%
  \makeatother%
  \begin{picture}(1,1.77380032)%
    \lineheight{1}%
    \setlength\tabcolsep{0pt}%
    \put(0,0){\includegraphics[width=\unitlength,page=1]{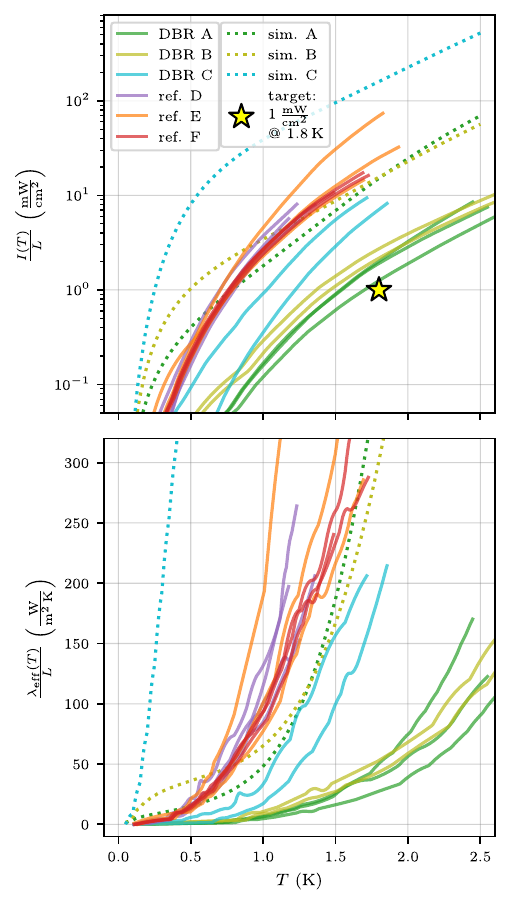}}%
    \put(0.02205593,1.73213653){\color[rgb]{0,0,0}\makebox(0,0)[lt]{\lineheight{1.25}\smash{\begin{tabular}[t]{l}\textbf{a}\end{tabular}}}}%
    \put(0.02164754,0.88741448){\color[rgb]{0,0,0}\makebox(0,0)[lt]{\lineheight{1.25}\smash{\begin{tabular}[t]{l}\textbf{b}\end{tabular}}}}%
  \end{picture}%
\endgroup%

    \caption{\label{fig:Fig4_Result}
    \textbf{Experimental results compared to simulations.}
    \textbf{a}~Iteratively determined heat integrals $I(T)$ for the DBR samples and reference structures toward $\SI{100}{mK}$ as a function of temperature. Multiple lines of the same color correspond to different samples of the same type, while dotted lines show the corresponding simulations. The $\SI{600}{nm}$ DBR (A) approaches the design target of $\SI{1}{mW/cm^2}$ at $\SI{1.8}{K}$.
    \textbf{b}~Effective thermal conductivity $\lambda_{\mathrm{eff}}(T)$ obtained from differentiation of the interpolated $I(T)$.
    }
\end{figure}

Figure~\ref{fig:Fig4_Result}\textcolor{crossrefBlue}{\textbf{a}} shows the heat integrals $I(T)$ obtained for the different DBR and reference samples, together with the corresponding simulations. Multiple curves of the same color correspond to measurements performed on different samples of the same type, illustrating the sample-to-sample variation. 
Note that to obtain the actual heat flux, $I(T_{\mathrm{L}})$ must be subtracted according to \Cref{eq:heat_flow}, however, for typical target temperatures, this results only in a small correction. 
Comparing the bulk reference sample E with the DBR sample B of similar total thickness demonstrates that the enhanced thermal insulation of the DBRs is not simply a thickness effect. Likewise, comparing the reference samples E and F shows that increasing the thickness of the \ce{SiO2} layer alone does not lead to improved thermal performance.

Differentiation of the interpolated heat integrals yields the effective thermal conductivity $\lambda_{\mathrm{eff}}(T)$ shown in Fig.~\ref{fig:Fig4_Result}\textcolor{crossrefBlue}{\textbf{b}}. The non-monotonic features originate from the differentiation of the interpolated data, which amplifies small experimental uncertainties. 

Outliers were removed only when necessary to ensure that the curve derived from the $\SI{100}{mK}$ data matches the corresponding curves obtained at higher $T_{\mathrm{MXC}}$, without artificial kinks arising from slight irregularities in the raw data, and the resulting oscillations in the derivative. For one sample each of types A and C and for all samples of type D, the last few data points measured at $\SI{100}{mK}$ were identified as outliers and excluded from the analysis. The same samples do not exhibit such deviations at elevated $T_{\mathrm{MXC}}$. Importantly, the heat integrals $I^{\SI{100}{mK}}(T)$ obtained from the reduced data sets remain consistent with the corresponding curves $I^{\geq \SI{200}{mK}}(T)$ derived from the full data sets, indicating that the exclusion does not affect the extracted thermal transport properties.

Figure \ref{fig:Fig4_Result} and \Cref{tab:Result} reveal a significant sample-to-sample variation. One potential concern is that this sample-to-sample spread might result from delamination of the DBR layers, which would improve thermal performance but likely make them unsuitable for via fabrication. No such delamination of the DBR stack was observed in TEM analysis (Appendix \ref{sec:SEM}) of select locations. 

%------------- Fig 6 Result ---------------
%--------ENDE Fig 6 Result----------------

\begin{table}[t]
    \centering
    \caption{\textbf{Exemplary sample-to-sample spread of $I(T)/L$.} The mean values and standard deviations of $I(T)/L$ for samples of the same type are computed at $\SI{1.0}{K}$, $\SI{1.5}{K}$, and $\SI{1.8}{K}$ using the curves displayed in \Cref{fig:Fig4_Result}\textcolor{crossrefBlue}{\textbf{a}}. All values are in mW/cm$^2$. Note that the values for DBRs A and B are one order of magnitude below that of the reference samples at $\SI{1.0}{K}$ and $\SI{1.5}{K}$ \cite{Denny_DBR}. For the references D, no data is available at $\SI{1.8}{K}$ as the thermal insulation property was not sufficient to reach $T_{\mathrm{H}}$. 
    The standard deviations for sample type E are large due to the spread between the curves. 
    Values without uncertainties indicate that only a single sample was available at the specified temperature.
    } 
    \label{tab:Result}
    \begin{ruledtabular}
        \begin{tabular}{ccccc}
           \shortstack{sample\\type} & \shortstack{\# meas. \\ samples} & \raisebox{0.3\normalbaselineskip}{$\SI{1.0}{K}$} & \raisebox{0.3\normalbaselineskip}{$\SI{1.5}{K}$} & \raisebox{0.3\normalbaselineskip}{$\SI{1.8}{K}$} \\
           \hline
           A & 3 & $0.150 \pm 0.026$& $0.80 \pm 0.16$ & $1.7  \pm 0.4$\\
           B & 2 & $0.25  \pm 0.06$ & $1.27 \pm 0.22$ & $2.4  \pm 0.4$\\
           C & 2 & $0.6  \pm 0.4$   & $4.2  \pm 1.9$  & $7$\\
           D & 4 & $2.9  \pm 0.6$   & -               & -\\
           E & 3 & $3.1  \pm 1.3$   & $18   \pm 12$   & $47  \pm 31$\\
           F & 3 & $2.60  \pm 0.16$ & $10.8 \pm 0.6$  & $17$\\
        \end{tabular}
    \end{ruledtabular}
\end{table}

The magnitude of the observed sample-to-sample variation is unexpectedly large as statistical fluctuations in the sample behavior seem unlikely, given their relatively large sample area of approximately $\SI{12.25}{mm^2}$. 

Possible reasons are layer thickness variations across a wafer and effects from the sample preparation, such as manual application of silver epoxy. 
While calculations in Appendix \ref{sec:influence_silver_epoxy_thermal_resistance} show that the thermal resistance of the silver epoxy is negligible, some epoxy squeezed out from the layer stack may cause a thermal short between the interposer and the sample substrate at the periphery of the sample (see \Cref{fig:Fig2_Setup_Samples}\textcolor{crossrefBlue}{\textbf{b}}.
Indeed, some measurements on samples where we took less care to avoid this effect showed reduced thermal isolation. This effect would lead to an underestimation of the isolation performance.

Despite this sample-to-sample variation, the DBRs consistently exhibit lower heat flow and, hence, better thermal insulation compared to both, the thinner DBRs and the reference samples.
For instance, a $\SI{600}{nm}$-thick Ta/\ce{SiO2} DBR A at $\SI{1.5}{K}$ reduces the heat flow by one order of magnitude compared to all reference samples of bare Si (sample D) or \ce{SiO2} (E, F), even when the latter more than two times thicker (F).

Most importantly, these findings confirm that the designed DBRs approach the thermal requirements for integrating cryogenic control electronics in close proximity to qubits. Specifically, a heat flow below about $\SI{1}{\milli W \per cm^2}$ is reached around $\SI{1.5}{K}$, close to the preferred $T_{\mathrm{H}}$ of $\SI{1.8}{K}$. The thickness of no more than $\SI{600}{nm}$ is compatible with high-density vias required for some large-scale spin-qubit architectures. According to simulations \cite{Denny_DBR}, thicker DBRs can even achieve significantly better performance well beyond the engineering target of $\SI{1}{\milli W \per cm^2}$ at $\SI{1.8}{K}$. Unfortunately, we could not yet explore such optimized structures due to  wafer bow resulting from the currently available fabrication process (see Appendix \ref{sec:growth}). Interestingly, $\SI{440}{nm}$- and $\SI{600}{nm}$-thick DBRs perform about one order of magnitude better than their simulations according to Ref.~\cite{Denny_DBR}.
A possible reason for the discrepancy is that phonon scattering other than due to the interfaces and variation of the layer thickness were not included in the simulation model. The comparison suggests that simulations based on the relatively simple model have limited quantitative predictive power for the present structures, capturing trends but not the absolute magnitude. 

Our results are comparable to the findings in Ref.~\cite{bon-mardion_superconducting_2024}, where lower powers and thus lower temperatures have been investigated in combination with superconducting, fully metallic DBRs in a periodic design. A direct comparison of the results is difficult because the temperature at the bottom of the stack has not been reported in that study, but has been set to $T_{\mathrm{L}} = T_{\mathrm{MXC}}$, whereas in our case, $T_{\mathrm{L}}$ is significantly higher.

\section{Conclusion \& Outlook} \label{sec:discussion_conclusion}

In summary, we present a concept for the integration of solid-state qubit chips with cryogenic control electronics that allows a substantial temperature gradient as well as a high density of microfabricated interconnects between the two parts, thus addressing one of the most significant scaling challenges. Our experiments demonstrate a very good isolation performance of a compact multilayer stack compatible with mechanically supported integration concepts. 
Compared to the widely propagated approach of operating semiconductor qubits at an elevated temperature of about $\SI{1}{K}$ \cite{vandersypen_hot_dense_coherent}, the solution presented here is compatible with lower qubit operating temperatures than elevated-temperatures operation, 
thus offering more leeway in system design trade-offs. 

While superconducting vias are not yet included in the present experiment, we estimate that thin films of disordered superconductors such as NbTiN would contribute only a negligible additional heat load, below $\SI{6}{\micro W/cm^2}$. 
This estimate is based on the via geometry depicted in \Cref{fig:Fig1_Vision} and obtained by calculating the electronic thermal conductivity from a typical normal-state resistivity of $\SI{1000}{\micro\ohm\centi\meter}$ \cite{TiN} using the Wiedemann--Franz law and accounting for the strong suppression of quasiparticle heat transport when operating well below the critical temperature $T_C$ (\textit{here:} $T/T_C = \SI{15}{\percent}$). 
The corresponding calculation is provided in Appendix~\ref{sec:Vias_Caluclation}. 

The concrete numbers used as target performance for our study were motivated by the needs for semiconductor qubits. It will be interesting to see what it takes to extend the concept to superconducting qubits, which are associated with more stringent temperature requirements and are more sensitive to electromagnetic excitations as well as nonequilibrium phonons. Additional electromagnetic filtering and shielding may help to address these additional challenges.

\section*{Acknowledgements}
This work was funded within the QSolid project by the German Federal Ministry of Research, Technology and Space (BMFTR) within the framework ``Quantum technologies – from basic research to market'' (Grant No. 13N16149). 

\section*{Data availability statement}
The data supporting the findings of this study is available
from the corresponding author upon reasonable request.

\section*{Competing interests}
H.B. is a co-inventor on a patent covering some of the content of the paper. The patent is licensed to ARQUE Systems GmbH, of which H.B. is a shareholder.
The other authors declare that they have no conflict of interest.

\newpage

\appendix %======================================================================================================

\section{Estimates used for our concrete concept of cryopackaging for close-to-qubit control} \label{sec:Estimates_concept}

Figure~\ref{fig:Fig1_Vision} illustrates our concrete concept of cryopackaging for close-to-qubit control that allows a higher power budget for the control electronics while maintaining the low qubit temperature, enabling compact and scalable integration in cryogenic environments.
To assess the dimensions and properties of the insulating matrix, we have to take the wiring density into account.
The wiring density is determined by the qubit spacing. This spacing is comfortably large for superconducting qubits, however, these qubits are potentially very sensitive to disturbance from electronics, such as high-frequency noise and infrared radiation, as well as dissipation near the qubits. Addressing these challenges is beyond the scope of this paper. 
In contrast, semiconductor qubits can be expected to have a more robust coherence, but their short native coupling range generally leads to small (i.e.~hundreds of $\si{\nano m}$) qubit spacing and thus a high wiring density \cite{SpinBus_Architecture}.

A concrete example for a quantum layer architecture based on semiconductor qubits with a qubit spacing compatible with our approach is the two-dimensional SpinBus architecture \cite{SpinBus_Architecture}, which is well-suited for implementing surface code error correction. 
This design is scalable at the quantum layer by tiling of a unit cell and reduces the qubit density by using shuttling lanes.
The variable length of these shuttling lanes can thus provide sufficient space for integrating cryogenic control electronics with the same footprint as the quantum layer unit cell being controlled. 
A unit cell size of $\SI{10}{\micro m}$ may reflect a good compromise between enough space for cryogenic electronics and avoiding excessively long shuttling distances, which would reduce the shuttling fidelity \cite{losert_oberlaender}. 4~million of these unit cells fit on a $\SI{2}{cm} \times \SI{2}{cm}$ chip. 
As a unit cell of the SpinBus architecture needs about 20 signal lines, leading to a required contact pitch of $\SI{2}{\micro m}$.
With a 10:1 aspect ratio feasible via standard etching such as the Bosch process \cite{Bosch} and $\leq \SI{50}{\percent}$ material removal for stability, the matrix thickness should thus be $\lesssim\SI{10}{\micro m}$.

To prevent electrical shorts, the sidewalls of the vias must be lined with a dielectric before deposition of a superconducting material. 
These superconducting vias are expected to have a low thermal conductivity at $T~<~T_C$ as electronic heat conduction is strongly suppressed in the superconducting state (see Appendix \ref{sec:Vias_Caluclation}). Electrical connection between cryogenic electronic chip and qubit chip can be established via flip-chip bonding with approx. $\SI{1}{\micro m}$ pitch, as state-of-the-art hybrid bonding reach a pitch as small as $\SI{0.5}{\micro m}$ \cite{flip_chip_0.5um_pitch,Hybrid_Bonding}.

Assuming a cooling power of $\SI{4}{\milli W}$ at $\SI{100}{mK}$, the power transmitted through the matrix must remain below around $\SI{1}{\milli W \per cm^2}$.  
Enabling the temperature of the electronics to be around $\SI{1.8}{K}$ thus requires a very low thermal conductance not achievable with a homogeneous material. For example, the transmitted power through $\SI{10}{\micro m}$ thin layers of graphite or Vespel SP-22 is above $\SI{100}{mW/cm^2}$ at $\SI{1.8}{K}$ \cite{LakeShoreAppendixI,Vespel}.

\section{Simulations of DBRs}
\label{sec:Simulations_LogLog}

In Ref.~\cite{Denny_DBR}, DBRs composed of silicon dioxide paired with tantalum, tungsten, or iridium are simulated. These materials are chosen because of their large mismatch in acoustic impedances (see \Cref{tab:Impedances_Simulation}).

\begin{table}[t]
    \centering
    \caption{\textbf{Overview of the materials simulated in Ref.~\cite{Denny_DBR} and their properties.} The acoustic impedance values for transverse polarization are taken from \cite{Denny_DBR}, where they are calculated via $Z_T = \varrho \cdot v_T$ with material density $\varrho$ and transverse speed of sound $v_T$. 
    } 
    \label{tab:Impedances_Simulation}
    \begin{ruledtabular}
        \begin{tabular}{cccc}
           material  & $\varrho\; (\si{kg \per (m^3)}$ & $v_T \; (\si{m \per s})$ & $Z \; (\si{kg \per (m^2 s)})$ \\
           \hline
            \ce{SiO2} & $\num{2.41E3}$ & $\num{3.53E3}$ & $\SI{0.85E7}{}$ \\
            \ce{Ta}   & $\num{1.66E4}$ & $\num{2.04E3}$ &$\SI{3.38E7}{}$  \\
            \ce{W}    & $\num{1.93E4}$ & $\num{2.74E3}$ &$\SI{5.29E7}{}$  \\
            \ce{Ir}   & $\num{2.25E4}$ & $\num{3.07E3}$ &$\SI{6.91E7}{}$  \\
        \end{tabular}
    \end{ruledtabular}
\end{table}

\begin{figure}
    \centering
    \includegraphics[width=1\linewidth]{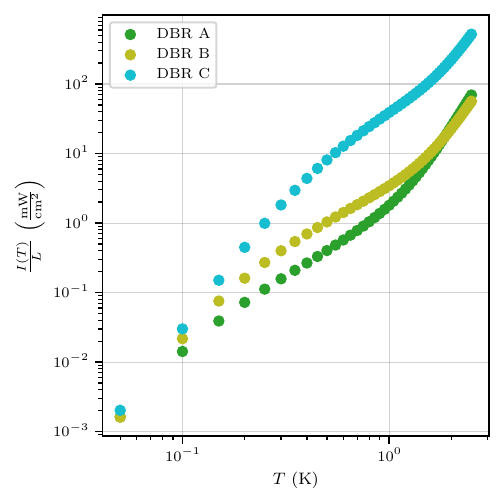}
    \caption{\label{fig:FigA_Simulations_LogLog}
    \textbf{Simulations of DBRs A, B, and C.} In log–log space, the slope between neighboring data points varies with $T$, making a power-law fit in linear representation across the entire range inappropriate.
    }
\end{figure}

In \Cref{fig:FigA_Simulations_LogLog}, the simulations for DBRs A, B, and C are displayed on a log–log scale. The progressions of the data points clearly show that global power-law fits are not an appropriate way to interpolate between the data points, as three areas with different slopes can be identified for each simulation.

Instead, we use PCHIP interpolation for the evaluation in \Cref{sec:Analysis}. Refer to \Cref{fig:Fig3_Analysis} for additional details on the analysis.

%------------- Fig 8 Simulations ---------------
%--------ENDE Fig 8 Simulations  ----------------

\section{Growth of DBRs} %================================================
\label{sec:growth}
Polycrystalline $\beta$-tantalum ($\beta$-Ta) thin films were deposited at room temperature via magnetron sputtering in a Ta-PVD Encore2\texttrademark{} chamber integrated into an Endura\texttrademark{} cluster tool (Applied Materials, Inc.). The deposition was performed using $\SI{5000}{W}$ DC power, $\SI{100}{W}$ AC bias power, and an Ar flow of 4 sccm at a chamber pressure of $\SI{0.065}{\pascal}$.

Amorphous silicon dioxide (\ce{SiO2}) thin films were deposited by plasma-enhanced atomic layer deposition (PEALD) in an EmerALD\textsuperscript{\raisebox{0.5ex}{\tiny\textregistered}} process module integrated into an ASM Eagle XP4 cluster system. The process utilized SAM24 -BDEAS-Bis(diethylamino)silane as the precursor and oxygen as the co-reactant, with a plasma power of $\SI{50}{W}$ and a substrate temperature of $\SI{323}{K}$.

\section{Setup} \label{subsec:Setup}

The experiments were performed in a dilution refrigerator (Oxford Triton 400) with a cooling power of approximately $\SI{300}{\micro \watt}$ at a MXC temperature of $\SI{100}{mK}$. The experimental setup is illustrated in \Cref{fig:Fig2_Setup_Samples}\textcolor{crossrefBlue}{\textbf{b}}. 
The setup consists of a sample (with or without a DBR or reference oxide) mounted face down on a silicon (Si) interposer. A commercial resistor (Panasonic ERA-6AEB6041V, 0805 package, $\SI{6.04}{k\ohm}$) was used as a Joule heater to emulate the power dissipation of a cryogenic control electronics chip. It was driven by a source meter (SMU Keithley 2450).

Several interfaces are present in the setup. While these do not change the final result due to $T_{\mathrm{L}}$ being measured and accounted for, their thermal impedance must be low enough to reach the desired regime and should thus not have a much higher thermal impedance than the device under test. 
For example, the thermal resistance (see Appendix \ref{sec:influence_silver_epoxy_thermal_resistance}) of gold in combination with molten indium foil at the Cu-Si interface is approximately $\SI{0.045}{(m^2 \cdot K)/W}$ at $\SI{1}{K}$ (see Appendix \ref{sec:Si-Cu-interface}), whereas that of a DBR sample is up to $\SI{0.21}{(m^2 \cdot K)/W}$ at the same temperature.
Starting from the MXC, the first interface is the Cu–Cu interface between the MXC and the sample holder. The holder is fabricated from oxygen-free high-conductivity (OFHC) copper and is tightly screwed onto the MXC to ensure good thermal contact. 
The typical thermal conductivity of OFHC copper (assuming residual-resistance ratio RRR = 100) at $\SI{0.5}{K}$ is approximately $\SI{100}{W/(m \cdot K)}$  \cite{OFHC_Cu} and is therefore not a limiting factor in this experiment given the holder length of approx. $\SI{10}{cm}$ and a cross section of approx. $\SI{6}{cm} \times \SI{0.5}{cm}$, leading to a low total estimated thermal impedance for our geometry of approx. $\SI{3.3}{mK/mW}$. To prevent oxidation and maintain low contact resistance, the holder surface was gold-plated.

The second interface is the Cu–Si contact between the sample holder and the Si interposer. The interposer was attached using a $\SI{50}{\micro m}$-thin molten indium foil, while the backside of the interposer was gold-plated to improve adhesion. 
Although this approach provides sufficient performance, it remains the primary thermal bottleneck in the cooling path, as the measured thermal conductance $C=\frac{\lambda_{\mathrm{eff}}\cdot A}{L}$ of the interface is approx. $\SI{1.2}{mW/(K)}$ at $\SI{0.5}{K}$ for the given sample dimensions (see also \Cref{fig:FigA_Interface2} in Appendix \ref{sec:Si-Cu-interface}). This thermal conductance corresponds to $\SI{30}{m}$ of high quality Cu and a thermal impedance of approx. $\SI{0.3}{m^2 K /W}$. 

The third interface is the Si–Si contact between the top of the interposer and the face of the sample. The sample (see \Cref{subsec:Samples}, \Cref{tab:Samples}, and \Cref{fig:Fig2_Setup_Samples}) was attached using silver epoxy (EPO-TEK\textsuperscript{\raisebox{0.5ex}{\tiny\textregistered}} H22), chosen for its ease of application in small quantities and its relatively good thermal performance (see also Appendix \ref{sec:Si-Cu-interface}). The same silver epoxy was used to bond the heater onto the top surface of the sample. Both Si pieces are intended to homogenize the temperature laterally, so that a uniform temperature is maintained even if the heat is applied locally.

Temperatures at the three positions ($T_{\mathrm{H}}$, $T_{\mathrm{L}}$, and $T_{\mathrm{MXC}}$) were measured using an AC resistance bridge (LakeShore Model 370) that recorded the resistances $R$ of negative-temperature-coefficient (NTC) resistors in a four-wire configuration to eliminate line resistance.
For $T_{\mathrm{MXC}}$, a calibrated ruthenium oxide sensor (LS-RX-102A-CD-0.05B, CD package) was mounted on the copper holder using brass screws to ensure good thermal contact. Using the manufacturer’s calibration curve, we converted the measured resistance $R_{\mathrm{MXC}}$ into $T_{\mathrm{MXC}}$ via linear interpolation in log–log space (see also \Cref{sec:Analysis} as well as Appendix \ref{sec:Measurement_Procedure} and \ref{sec:ThermometryMethods} for details).
For $T_{\mathrm{H}}$ and $T_{\mathrm{L}}$, i.e. the temperatures at the top and bottom of the sample, we use uncalibrated Cernox CX-1010-BR-HT sensors because of size and cost considerations. The sensors were glued onto the sample and interposer, respectively, with GE varnish so that the sensors can be reused.
Both sensor types are suitable down to $\SI{50}{mK}$ and, owing to their NTC characteristics, exhibit increasing resistance and its derivative, and thus sensitivity, towards $\SI{0}{K}$. 

The wiring employed highly resistive CuNi DC lines (approximately $\SI{130}{\ohm}$ at $\SI{100}{mK}$ between the breakout box and the MXC), which were thermally anchored at each stage of the dilution refrigerator. The lines are connected to a PCB that features a cutout for the copper holder. This cutout allows us to directly place the interposer on the copper holder to improve the thermal contact at the Cu–Si interface. Manually placed wire bonds (Al ($\SI{1}{\percent}$ Si), $\text{\diameter} = \SI{25}{\micro m}$) connected the PCB to the interposer and sample (see also Appendix~\ref{sec:ThermometryMethods}).

\section{Optimization of the Cu-Si interface} \label{sec:Si-Cu-interface}

%------------- Fig 8 ---------------
\begin{figure}
    \centering
    \def\svgwidth{3.4in}
    %% Creator: Inkscape 1.2.2 (b0a8486, 2022-12-01), www.inkscape.org
%% Accompanies image file 'Fig8_Si-Cu_abc.pdf' (pdf, eps, ps)
\begingroup%
  \makeatletter%
  \providecommand\color[2][]{%
    \errmessage{(Inkscape) Color is used for the text in Inkscape, but the package 'color.sty' is not loaded}%
    \renewcommand\color[2][]{}%
  }%
  \providecommand\transparent[1]{%
    \errmessage{(Inkscape) Transparency is used (non-zero) for the text in Inkscape, but the package 'transparent.sty' is not loaded}%
    \renewcommand\transparent[1]{}%
  }%
  \providecommand\rotatebox[2]{#2}%
  \newcommand*\fsize{\dimexpr\f@size pt\relax}%
  \newcommand*\lineheight[1]{\fontsize{\fsize}{#1\fsize}\selectfont}%
  \ifx\svgwidth\undefined%
    \setlength{\unitlength}{244.80000173bp}%
    \ifx\svgscale\undefined%
      \relax%
    \else%
      \setlength{\unitlength}{\unitlength * \real{\svgscale}}%
    \fi%
  \else%
    \setlength{\unitlength}{\svgwidth}%
  \fi%
  \global\let\svgwidth\undefined%
  \global\let\svgscale\undefined%
  \makeatother%
  \begin{picture}(1,1.40015083)%
    \lineheight{1}%
    \setlength\tabcolsep{0pt}%
    \put(0,0){\includegraphics[width=\unitlength,page=1]{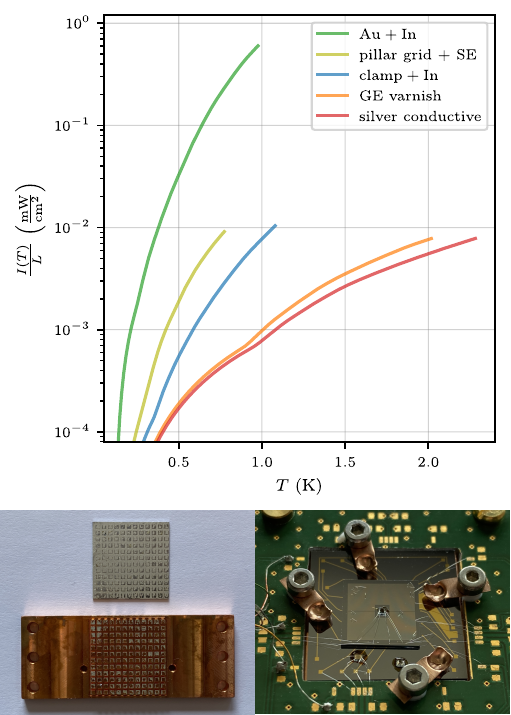}}%
    \put(0.02205594,1.35848705){\color[rgb]{0,0,0}\makebox(0,0)[lt]{\lineheight{1.25}\smash{\begin{tabular}[t]{l}\textbf{a}\end{tabular}}}}%
    \put(0.02164743,0.34864226){\color[rgb]{0,0,0}\makebox(0,0)[lt]{\lineheight{1.25}\smash{\begin{tabular}[t]{l}\textbf{b}\end{tabular}}}}%
    \put(0.52160658,0.34864226){\color[rgb]{1,1,1}\makebox(0,0)[lt]{\lineheight{1.25}\smash{\begin{tabular}[t]{l}\textbf{c}\end{tabular}}}}%
    \put(0,0){\includegraphics[width=\unitlength,page=2]{Fig8_Si-Cu_abc.pdf}}%
    \put(0.35732849,0.29131852){\color[rgb]{0,0,0}\makebox(0,0)[lt]{\lineheight{1.25}\smash{\begin{tabular}[t]{l}20 mm\end{tabular}}}}%
    \put(0,0){\includegraphics[width=\unitlength,page=3]{Fig8_Si-Cu_abc.pdf}}%
    \put(0.74511082,0.29131852){\color[rgb]{1,1,1}\makebox(0,0)[lt]{\lineheight{1.25}\smash{\begin{tabular}[t]{l}20 mm\end{tabular}}}}%
  \end{picture}%
\endgroup%

    \caption{\label{fig:FigA_Interface2}
    \textbf{Comparison of heat integrals towards $\SI{100}{mK}$ of different materials at the Cu-Si-interface.} 
    \textbf{a} The five interface solutions tested show large variations in thermal performance, resulting in different surface temperatures of the Si layers for identical applied power. Molten indium between gold-plated Cu and Si provides the highest thermal conductance and was therefore selected as the interface material for all subsequent measurements. While adequate for the thermal conductivity experiments at the Si–Si interface, this configuration is not yet optimal and leaves room for improvement.
    \textbf{b} The pillar grid holder has $\SI{1.5}{mm}$ deep trenches and is used in combination with silver epoxy (SE).
    \textbf{c} Four copper beryllium clamps are used to press the interposer on non-molten indium foil.
    Note that the sample used for this test (inner square) is approx. 8 times larger than the samples analyzed in \Cref{sec:Analysis}.
    }
\end{figure}
%--------ENDE Fig 8----------------

Although GE varnish is commonly used for gluing Si samples to either silicon or copper in cryogenic applications \cite{GEVarnish_LakeShore, GEVarnish}, this type of adhesive is not ideal in terms of thermal conductivity.
Figure~\ref{fig:FigA_Interface2} compares five different interface solutions tested for joining silicon to copper: silver conductive, GE varnish, a clamp holder with $\SI{125}{\micro m}$ thin indium foil, a grid pillar holder with silver epoxy (SE), and indium foil molten between gold-plated surfaces. The molten indium foil was initially $\SI{200}{\micro m}$ thin, 
but for indium foils thicker than $\SI{50}{\micro m}$, excess indium tends to flow to the sides of the interposer during pressing and melting, and then solidifies into small beads, which can be easily removed after cooling. 
Among these, indium molten between a gold-plated interposer backside and a gold-plated copper holder provided the best thermal contact. The Si interposer was directly attached to the Cu sample plate, which was screwed onto the holder mounted on the MXC. Improving this Cu-Si interface was a prerequisite for enabling precise measurements of the subsequent Si–Si interface. While this configuration was sufficient for the present measurements, evidenced by a clear temperature drop across the Si–Si interface, it still offers potential for further optimization.

\section{Measurement procedure}
\label{sec:Measurement_Procedure}
During the measurements, a PID control loop stabilized the mixing chamber temperature $T_{\mathrm{MXC}}$ at $\SI{100}{mK}$ to $\SI{700}{mK}$ in steps of $\SI{100}{mK}$, provided that the heating power applied remained below the available cooling power of the dilution refrigerator. 
We repeated the experiments at different mixing chamber temperatures for two reasons: 
First, measurements without on-chip heater power $P$ applied enable post-calibration of the uncalibrated temperature sensors. For this purpose, we conducted additional calibration measurements at elevated temperatures of $\SI{800}{mK}$, $\SI{900}{mK}$, $\SI{1.3}{K}$, $\SI{1.7}{K}$, and $\SI{2.2}{K}$. The upper limit of $\SI{2.2}{K}$ corresponds to the maximum temperature that can be stably maintained by the PID loop.
Second, stepping the heater power $P$ from $\SI{0}{W}$ to $\SI{1}{mW}$ at every temperature setpoint $\leq \SI{700}{mK}$ provides redundant data and allows us to later verify the consistency of the analysis and results (see \Cref{sec:Analysis}).

After adjusting the heater power, the system was allowed to equilibrate for $\SI{1}{min}$ to ensure steady-state conditions. For each power value, 10 readings were averaged. 
Following each change in $T_{\mathrm{MXC}}$, we allowed for a wait time of $\SI{5}{min}$ to reach a steady state. We confirmed temperature stability before each new measurement sequence: After the specified waiting times, the sensor's resistances and thus temperatures do no longer change. Thus, we assume thermal equilibrium, i.e. $T_{\mathrm{H}} = T_{\mathrm{L}} = T_{\mathrm{MXC}}$, allowing calibration of the uncalibrated sensors at zero heating power ($P = \SI{0}{mW}$). The corresponding resistances $R_{\mathrm{H}}$ and $R_{\mathrm{L}}$ were mapped to $T_{\mathrm{H}} = T_{\mathrm{MXC}}$ and $T_{\mathrm{L}} = T_{\mathrm{MXC}}$, respectively. The resulting calibration curves for the two Cernox sensors were generated via linear interpolation in log–log space between the calibration points. Subsequently, for $P > \SI{0}{W}$, we converted the measured $R_{\mathrm{H}}$, $R_{\mathrm{L}}$, and  $R_{\mathrm{MXC}}$ values to $T_{\mathrm{H}}$, ${T_{\mathrm{L}}}$, and ${T_{\mathrm{MXC}}}$ using these interpolated calibration functions. 

\section{Thermometry errors} \label{sec:ThermometryMethods}
In this section, we discuss four potential sources of error in the temperature readings: 
\textit{(i)} lateral positioning of the sensors, 
\textit{(ii)} thermal contact of the sensors (including additional interfaces and heat-sinking shunts via wire bonds), 
\textit{(iii)} in-house calibration of uncalibrated sensors using calibrated references, and 
\textit{(iv)} self-heating of the sensors during measurement. 
Where applicable, we explicitly indicate whether a given error leads to an over- or underestimation of the thermal insulation capability. 
Estimates for all contributions are summarized in \Cref{tab:Error_Estimates}.

\paragraph{Lateral positioning of the sensors.}
Due to geometric constraints, it was not possible to align the heater and temperature sensors perfectly on top of each other. Instead, there is a lateral displacement $L$ of approximately $\SI{2}{mm}$ for the $T_{\mathrm{H}}$ sensor and $\SI{7}{mm}$ for the $T_{\mathrm{L}}$ sensor from the heater position. This displacement leads to a small lateral temperature gradient in the silicon pieces and thus to a slight underestimation of both temperatures.

Silicon exhibits a strongly temperature-dependent thermal conductivity $\lambda(T)$ ranging from approximately $\SI{11}{W/(m \cdot K)}$ to $\SI{2700}{W/(m \cdot K)}$ between $\SI{290}{mK}$ and $\SI{1.8}{K}$ \cite{Si_thermal_conductivity}. 
To estimate an upper bound of the resulting temperature drop, we calculate the lateral thermal resistance as
\begin{equation}
    R_{\mathrm{th}} = \frac{L}{\lambda(T)\cdot A},
\end{equation}
where the cross-sectional area $A$ is given by the wafer thickness of $\SI{750}{\micro m}$ and the lateral dimensions of the Si pieces. This estimate corresponds to the case where the full heat flux flows laterally through the Si over the distance $L$ and should thus reflect a conservative upper bound.
Using the measured temperatures for all applied heater powers (see raw data in \Cref{fig:FigB_Rawdata_2}), we find that the largest gradient occurs for $T_{\mathrm{L}}$ of the DBR sample, where the lowest temperatures, and thus the lowest silicon thermal conductivities, are reached for a given power.

For all applied powers, the estimated temperature difference remains below $\SI{0.3}{mK}$, corresponding to less than $\SI{0.1}{\percent}$, and is therefore negligible. Due to the high thermal conductivity of OFHC copper, the larger lateral displacement of the holder sensor (approx. $\SI{2}{cm}$) can likewise be neglected.

\paragraph{Thermal contact of the sensors.}
The finite thermal contact of the sensors to the temperature to be measured 
in conjunction with additional heat sinking via the PCB and bond wires to the MXC lead to an underestimation of the measured temperatures, i.e. $T_{\mathrm{H}}$ and $T_{\mathrm{L}}$. 

While underestimating $T_{\mathrm{L}}$ would in principle overestimate the thermal isolation, this effect is weak because, for the large temperature differences considered here, the heat flow depends much more strongly on $T_{\mathrm{H}}$ than on $T_{\mathrm{L}}$. In contrast, underestimating $T_{\mathrm{H}}$ leads to an underestimation of the thermal insulation capability and therefore constitutes a conservative, or \textit{safe}, error. Consequently, the actual thermal insulation performance of the samples may be better than inferred from the measured data. This effect increases with temperature and therefore affects the DBR samples more strongly than the reference structures. The fact that varying  $T_{\mathrm{MXC}}$ between $\SI{100}{mK}$ and $\SI{700}{mK}$ does not alter the results for $I(T)$ indicates that this effect is not very large.

The heat flux through the PCB and the bond-wire shunts is negligible compared with the applied heater power and therefore does not significantly affect the heat flow through the sample interface used in the analysis. 
The measurement setup provides two parallel heat-flow paths: one through the sample interface and one through four aluminum bond wires connecting the sample to the PCB and hence to the MXC. Each bond wire has a diameter of $\SI{25}{\micro m}$ and a length of approximately $\SI{7.5}{mm}$. The total applied power can thus be written as
\begin{equation}
    P = P_{\mathrm{interface}} + P_{\mathrm{W}},
\end{equation}
where $P_{\mathrm{W}}$ denotes the power dissipated through the bond wires.

The thermal conductivity of the wire bond material used, i.e. aluminum bond wires alloyed with $\SI{1}{\percent}$ silicon, was measured as $\SI{0.001}{W/(m \cdot K)}$ at $\SI{100}{mK}$ and $\SI{1}{W/(m \cdot K)}$ at $\SI{1}{K}$ in \cite{Wirebonds}.
Thus, $P_{\mathrm{W}}$ can be estimated from the heat flow as
\begin{equation}
    P_{\mathrm{W}} = 4 \cdot \frac{A_{\mathrm{W}}}{L_{\mathrm{W}}} \cdot \int_{\SI{0.1}{K}}^{\SI{1}{K}} \lambda(T) dT \approx \SI{0.1}{\micro W}
\end{equation}

where $\lambda(T)$ denotes the thermal conductivity of the wire bond material, taken from \cite{Wirebonds}.

In our experiments at $\SI{100}{mK}$, the top-side temperature $T_{\mathrm{H}}$ typically reaches about $\SI{1}{K}$ for applied powers in the range of $\SI{10}{\micro W} \leq P \leq \SI{50}{\micro W}$, depending on the sample (see also the raw data in \Cref{fig:FigB_Rawdata_2}). Accordingly, the power dissipated through the bond wires accounts for only $\SI{1}{\percent}$ to $\SI{0.2}{\percent}$ of the total applied power and is therefore negligible.

For $P=\SI{1}{mW}$, $T_{\mathrm{L}}$ and $T_{\mathrm{H}}$ reach temperatures up to $\SI{1.9}{K}$ and $\SI{2.9}{K}$, respectively (see \Cref{fig:FigB_Rawdata_2}). At these temperatures, the aluminium bond wires are no longer superconducting. We calculate the $P_W$ based on estimated data for Al 1200 \cite{Al_wire_1K-to-4K}:
\begin{align}
P_{\mathrm{W, L}} &= 4 \cdot \frac{A_{\mathrm{W}}}{L_{\mathrm{W}}} \cdot \int_{\SI{1}{K}}^{\SI{1.9}{K}} \lambda(T) dT \approx \SI{3}{\micro W}
\\
P_{\mathrm{W, H}} &= 4 \cdot \frac{A_{\mathrm{W}}}{L_{\mathrm{W}}} \cdot \int_{\SI{1}{K}}^{\SI{2.9}{K}} \lambda(T) dT \approx \SI{12}{\micro W}
\end{align}

These values correspond to $\SI{0.3}{\percent}$ and $\SI{1.2}{\percent}$ of the power applied. 

\paragraph{In-house calibration of the Cernox sensors.}
A further potential error source arises from the in-house calibration of the Cernox sensors using a calibrated ruthenium oxide sensor and linear interpolation in log-log space (see \Cref{subsec:Setup}). For $T \leq \SI{4.2}{K}$, the manufacturer’s calibration certificate of the ruthenium oxide sensor specifies expanded uncertainties of $\pm \SI{4}{mK}$ with a coverage factor $k=2$, corresponding to a confidence level of $\SI{95}{\percent}$.

The in-house calibration relies on twelve calibration points between $\SI{100}{mK}$ and $\SI{2.2}{K}$, compared to the 30 points provided by the manufacturer. Increasing the number of calibration points did not lead to a significant change of the calibration curve and was therefore not pursued further. To quantify the associated interpolation error, we performed a leave-one-out cross validation (LOOCV): one calibration point $\{R_i(P=\SI{0}{W}), T_i(P=\SI{0}{W})\}$ was removed prior to generating the calibration curve, which was then used to determine the temperature $T$ corresponding to $R_i(P=\SI{0}{W})$. The resulting deviation $\Delta T = T - T_i(P=\SI{0}{W})$ remains below $\SI{3}{\percent}$ for the calibration points $i \in [3, 11]$.

For large applied powers on DBR samples, the calibration curve for $T_{\mathrm{H}}$ must sometimes be extrapolated beyond $\SI{2.2}{K}$, which corresponds to the maximum temperature of the PID loop. As described in \Cref{sec:Measurement_Procedure}, the extrapolation follows the slope defined by the last two calibration points in log-log space. 
This ignores the expected curvature (steeper slope in plot $T$ vs.~$R$), so that $T_{\mathrm{H}}$ is underestimated, leading to a conservative estimate.
To quantify the underestimation, we assumed an increase of the slope by $\SI{10}{\percent}$, resulting in a worst-case deviation of ca.~$\SI{200}{mK}$ for the smallest measured resistances. 
Overall, uncertainties related to interpolation and extrapolation of the calibration curves remain small or act in the conservative direction and do not alter the overall conclusions.

\paragraph{Self-heating of the sensors.}
A voltage excitation of $\SI{63.2}{\micro V}$ was used for $T_{\mathrm{MXC}}~<~\SI{500}{mK}$ and increased to $\SI{200}{\micro V}$ at higher temperatures. This corresponds to excitation powers below $\SI{2.3}{pW}$ and results in a self-heating error of less than $\SI{0.3}{mK}$ \cite{Lakeshore_370_Manual}, which is negligible for the present analysis.

Taken together, these estimates demonstrate that thermometry-related uncertainties cannot account for the observed differences between the DBR samples and the reference structures, indicating that the improved performance of the DBR samples reflects a genuine physical effect.

\begin{table}[t]
    \centering
    \caption{\textbf{Estimates of potential error sources.} Safe direction means here that the thermal insulation capability of the DBRs is more strongly underestimated than that of the references.} 
    \label{tab:Error_Estimates}
    \begin{ruledtabular}
        \begin{tabular}{cc}
           potential error source  & estimate\\
           \hline
           lateral positioning (through Si)     & $<~\SI{0.3}{mK}$ \\
           thermal contact of sensors           & \parbox{3cm}{safe direction} \\
           in-house calibration                 & $<~\SI{5}{\percent}$   \\
           self-heating of sensors              & $<~\SI{0.3}{mK}$ \\
        \end{tabular}
    \end{ruledtabular}
\end{table}

\section{TEM image of DBR A} \label{sec:SEM}

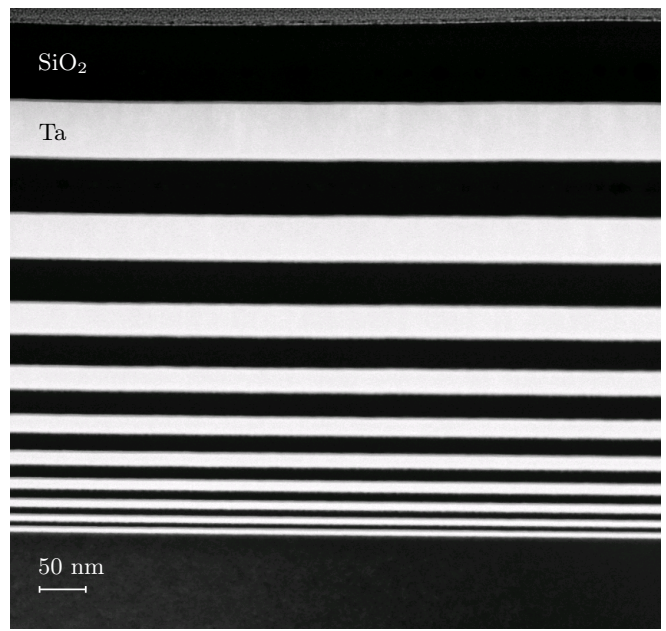
\begin{figure}
    \centering
    \def\svgwidth{3.4in}
    %% Creator: Inkscape 1.2.2 (b0a8486, 2022-12-01), www.inkscape.org
%% Accompanies image file 'Fig9_TEM.pdf' (pdf, eps, ps)
\begingroup%
  \makeatletter%
  \providecommand\color[2][]{%
    \errmessage{(Inkscape) Color is used for the text in Inkscape, but the package 'color.sty' is not loaded}%
    \renewcommand\color[2][]{}%
  }%
  \providecommand\transparent[1]{%
    \errmessage{(Inkscape) Transparency is used (non-zero) for the text in Inkscape, but the package 'transparent.sty' is not loaded}%
    \renewcommand\transparent[1]{}%
  }%
  \providecommand\rotatebox[2]{#2}%
  \newcommand*\fsize{\dimexpr\f@size pt\relax}%
  \newcommand*\lineheight[1]{\fontsize{\fsize}{#1\fsize}\selectfont}%
  \ifx\svgwidth\undefined%
    \setlength{\unitlength}{635.71029543bp}%
    \ifx\svgscale\undefined%
      \relax%
    \else%
      \setlength{\unitlength}{\unitlength * \real{\svgscale}}%
    \fi%
  \else%
    \setlength{\unitlength}{\svgwidth}%
  \fi%
  \global\let\svgwidth\undefined%
  \global\let\svgscale\undefined%
  \makeatother%
  \begin{picture}(1,0.95588314)%
    \lineheight{1}%
    \setlength\tabcolsep{0pt}%
    \put(0,0){\includegraphics[width=\unitlength,page=1]{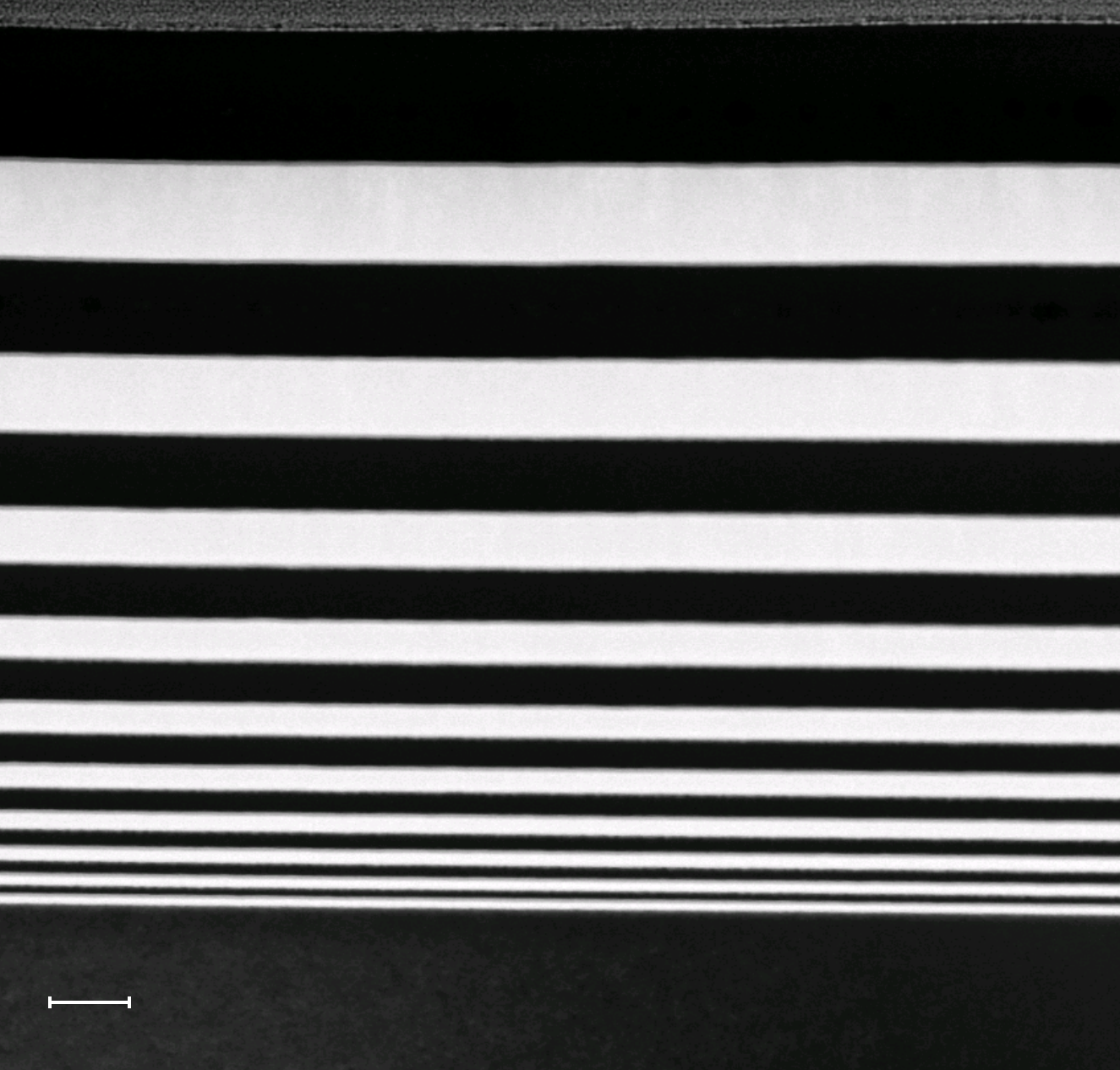}}%
    \put(0.04247708,0.860682){\color[rgb]{1,1,1}\makebox(0,0)[lt]{\lineheight{1.25}\smash{\begin{tabular}[t]{l}SiO$_2$\end{tabular}}}}%
    \put(0.04323214,0.75196243){\color[rgb]{1,1,1}\makebox(0,0)[lt]{\lineheight{1.25}\smash{\begin{tabular}[t]{l}\textcolor{black}{Ta}\end{tabular}}}}%
    \put(0.04247708,0.08449516){\color[rgb]{1,1,1}\makebox(0,0)[lt]{\lineheight{1.25}\smash{\begin{tabular}[t]{l}50 nm\end{tabular}}}}%
  \end{picture}%
\endgroup%

    \caption{\label{fig:FigC_SEM}
    \textbf{Transmission electron microscopy (TEM) high-angle annular dark-field (HAADF) cross-sectional image of a sample of type A before cooldown.}  The \ce{SiO2} layers appear dark, whereas the Ta layers appear white.
    The multilayer structure appears flat and well-defined, with no evidence of delamination or voids.
    }
\end{figure}

The transmission electron microscopy (TEM) high-angle annular dark-field (HAADF) cross-sectional image of a sample of type A shown in \Cref{fig:FigC_SEM} was taken prior to thermal measurements to investigate whether delamination could explain the variation observed among samples from the same wafer (see also \Cref{sec:Results}). The cross section reveals uniform and continuous layering, without visible voids or irregularities. Therefore, delamination is unlikely to be the source of the experimental spread observed for this sample type, unless delamination occurs only during cooldown.

%------------- Fig 9 ---------------
%--------ENDE Fig 9 ----------------

\section{Influence of silver epoxy on $\lambda(T)$} \label{sec:influence_silver_epoxy_thermal_resistance}

To estimate the influence of the silver epoxy used to attach the samples to the Si interposer on the measured results, we evaluate the thermal impedance $Z_{\mathrm{th}}$, defined as the inverse of the \textit{effective} thermal conductivity per thickness $L$:
\begin{equation}
    Z_{\mathrm{th}} = \frac{L}{\lambda_{\mathrm{eff}}(T)} \medspace .
\end{equation}

For stacked layers, the total thermal impedance is the sum of the individual contributions:
\begin{equation}
    Z_{\mathrm{total}}(T) = Z_{\mathrm{DBR}}(T) + Z_{{\mathrm{SE}}}(T) + Z_{\ce{Si}}(T) \medspace .
\end{equation}

The contribution of the silicon substrate, i.e. $Z_{\ce{Si}}(T)$, is negligible in the relevant temperature range, i.e. above $\SI{1}{K}$. At $\SI{1.5}{K}$ \cite{Si_thermal_conductivity}, for example, the contribution can be calculated as 
\begin{equation}
    Z_{\ce{Si}}(\SI{1.5}{K}) = \frac{\SI{750}{\micro m}}{\SI{16.2}{W/(cm \cdot K)} } \approx \SI{5E-7}{m^2 K / W} \medspace .
\end{equation}

Accordingly, the thermal impedance of silver epoxy can be derived directly from the effective thermal conductivities of the reference samples at $\SI{1.5}{K}$:
\begin{equation}
    Z_{\mathrm{SE}} = \SI{0.004}{m^2 K / W}  \medspace .
\end{equation}

The total thermal impedance of the DBR A sample is approximately 
\begin{equation}
    Z_{\mathrm{DBR \thinspace A, total}} = \SI{0.05}{m^2 K / W} 
\end{equation}

which clearly shows that
\begin{equation}
    Z_{\mathrm{SE}}(T) \ll Z_{\mathrm{total}}(T) \quad \xrightarrow{} \quad \lambda_{\mathrm{DBR}}(T) \approx \lambda_{\mathrm{total}} \medspace .
\end{equation} 

\begin{figure}[t]
    \centering
    \def\svgwidth{3.4in}
    %% Creator: Inkscape 1.2.2 (b0a8486, 2022-12-01), www.inkscape.org
%% Accompanies image file 'Fig10_Rth.pdf' (pdf, eps, ps)
\begingroup%
  \makeatletter%
  \providecommand\color[2][]{%
    \errmessage{(Inkscape) Color is used for the text in Inkscape, but the package 'color.sty' is not loaded}%
    \renewcommand\color[2][]{}%
  }%
  \providecommand\transparent[1]{%
    \errmessage{(Inkscape) Transparency is used (non-zero) for the text in Inkscape, but the package 'transparent.sty' is not loaded}%
    \renewcommand\transparent[1]{}%
  }%
  \providecommand\rotatebox[2]{#2}%
  \newcommand*\fsize{\dimexpr\f@size pt\relax}%
  \newcommand*\lineheight[1]{\fontsize{\fsize}{#1\fsize}\selectfont}%
  \ifx\svgwidth\undefined%
    \setlength{\unitlength}{244.79999734bp}%
    \ifx\svgscale\undefined%
      \relax%
    \else%
      \setlength{\unitlength}{\unitlength * \real{\svgscale}}%
    \fi%
  \else%
    \setlength{\unitlength}{\svgwidth}%
  \fi%
  \global\let\svgwidth\undefined%
  \global\let\svgscale\undefined%
  \makeatother%
  \begin{picture}(1,1.75294566)%
    \lineheight{1}%
    \setlength\tabcolsep{0pt}%
    \put(0,0){\includegraphics[width=\unitlength,page=1]{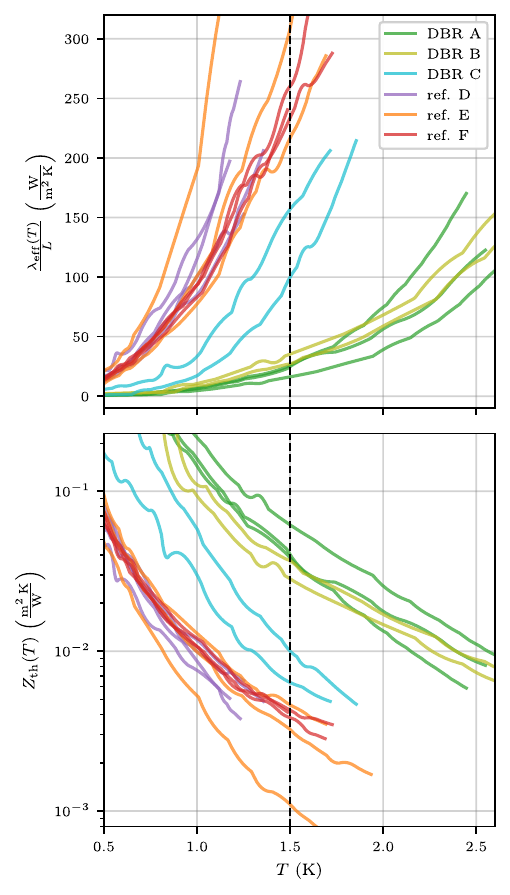}}%
    \put(0.02205593,1.71128187){\color[rgb]{0,0,0}\makebox(0,0)[lt]{\lineheight{1.25}\smash{\begin{tabular}[t]{l}\textbf{a}\end{tabular}}}}%
    \put(0.02164754,0.87582338){\color[rgb]{0,0,0}\makebox(0,0)[lt]{\lineheight{1.25}\smash{\begin{tabular}[t]{l}\textbf{b}\end{tabular}}}}%
  \end{picture}%
\endgroup%

    \caption{\label{fig:FigD_R_th}
    \textbf{Calculation of thermal resistances.} 
    Calculated thermal resistances show that the contribution of the silver epoxy to the total effective thermal conductivity of the DBR samples is negligible at the relevant temperatures, e.g., at $\SI{1.5}{K}$ (dashed black line).
    }
\end{figure}

Figure~\ref{fig:FigD_R_th} illustrates that the contribution of silver epoxy to the overall thermal resistance is negligible. Therefore, the spread mentioned in \Cref{sec:Results} cannot be attributed to varying thickness of the silver epoxy.

%------------- Fig 10 R_th ---------------
%--------ENDE Fig 10 R_th----------------

\section{Heat flow through superconducting vias}
\label{sec:Vias_Caluclation}

To estimate the heat load introduced by superconducting vias, we consider a cylindrical shell geometry with outer diameter of $\SI{1}{\micro\meter}$, shell thickness of $\SI{20}{nm}$, pitch $p=\SI{2}{\micro\meter}$, and length $l=\SI{10}{\micro\meter}$ (see \Cref{fig:Fig1_Vision}). The cross-sectional area of the superconducting shell is $A_\mathrm{via} \approx \SI{6.2e-14}{m^2}$.
With one via per area $p^2$, this corresponds to a metal fill factor $f = \frac{A_\mathrm{via}}{p^2}\approx 1.5\times 10^{-2}$.

Assuming a typical normal-state resistivity of $\varrho_\mathrm{n}~=~\SI{1000}{\micro\ohm\centi\meter}$ 
\cite{TiN}, the electronic thermal conductivity in the normal state is estimated from the Wiedemann--Franz law \cite{AshcroftMermin1976},
\begin{equation}
    \lambda_\mathrm{n}(T) = \frac{L_0 T}{\varrho_\mathrm{n}},
\end{equation}
where $L_0=\SI{2.44e-8}{W\ohm/K^2}$ is the Lorenz number. The areal heat flux through the via array is then
\begin{equation}
    Q_\mathrm{n} = \frac{f}{l}\int_{T_\mathrm{L}}^{T_\mathrm{H}} \lambda_\mathrm{n}(T)\,\mathrm{d}T
    = \frac{f}{l}\frac{L_0}{2\varrho_\mathrm{n}}\left(T_\mathrm{H}^2-T_\mathrm{L}^2\right).
\end{equation}
For $T_\mathrm{H}=\SI{1.8}{K}$ and $T_\mathrm{L}=\SI{100}{mK}$, this gives
\begin{equation}
    Q_\mathrm{n} \approx \SI{0.6}{mW/cm^2}  .
\end{equation}

Below the superconducting transition temperature $T_C$, the electronic thermal conductivity is strongly reduced because the quasiparticle population is suppressed \cite{Bardeen}. With $T_C(\mathrm{NbTiN}) \approx\SI{12}{K}$ \cite{Telkamp2025}, operation at $\SI{1.8}{K}$ corresponds to $T/T_C = 0.15$. Assuming conservatively that the heat transport is reduced by a factor 1/100, the superconducting-state heat flux becomes
\begin{equation}
    Q_\mathrm{SC} \approx \SI{6}{\micro W/cm^2}.
\end{equation}
This value is well below the target heat-load budget of $\SI{1}{mW/cm^2}$ and therefore supports the assumption discussed in \Cref{sec:discussion_conclusion} that superconducting vias add only a negligible thermal load in the present cryopackaging concept.

\newpage
\bibliography{bibliography}

\end{document}